\documentclass[singlecolumn]{article}

\usepackage{arxiv}

\usepackage[utf8]{inputenc} 
\usepackage[T1]{fontenc}    
\usepackage{hyperref}       
\usepackage{url}            
\usepackage{booktabs}       
\usepackage{array}          
\usepackage{amsfonts}       
\usepackage{nicefrac}       
\usepackage{microtype}      
\usepackage{graphicx}
\graphicspath{{figures/}} 
\usepackage{enumitem}       
\usepackage[authoryear,round,sort&compress]{natbib}
\setcitestyle{authoryear,round,citesep={;},aysep={,},yysep={,}}
\usepackage{doi}

\usepackage{math-notation}         
\usepackage{cleveref}       

\title{
Thermodynamically consistent neural constitutive models for elastoplasticity via eigen-representations
}

\author{Harikrishnan Vijayakumaran\\
	Faculty of Mechanical Engineering\\
	Delft University of Technology\\
	Delft, 2628 CD\\
	The Netherlands\\
	\And
	Shunyu Yin\\
	School of Engineering\\
	Brown University\\
	Providence, RI 02912\\
	United States of America\\
    \And
    Martin Abendroth \\
	Institute of Mechanics and Fluid Dynamics\\
	TU Bergakademie Freiberg\\
	09599 Freiberg\\
	Germany \\
	\And
	Miguel A. Bessa\thanks{Corresponding author, email: \texttt{miguel\_bessa@brown.edu}.} \\
	School of Engineering\\
	Brown University\\
	Providence, RI 02912\\
	United States of America\\
	\And
    Karl A. Kalina\thanks{Corresponding author, email: \texttt{karl.kalina@tu-dresden.de}.} \\ 
	Institute of Solid Mechanics\\
	TU Dresden\\
	01062 Dresden\\
	Germany \\
}

\renewcommand{\shorttitle}{Thermodynamically consistent path-dependent neural constitutive models}

\usepackage{pifont}
\usepackage{float}  

\usepackage{amssymb}

\usepackage{amsthm}
\usepackage{amsmath}
\usepackage{siunitx}  
\usepackage{mathtools}		
\mathtoolsset{centercolon}	
\usepackage{newtxmath}      
\usepackage{lineno}
\usepackage{bbm}
\usepackage[dvipsnames]{xcolor}
\usepackage{multirow}

\usepackage{epstopdf}
\usepackage[normalem]{ulem}
\usepackage{soul}          

\definecolor{colHV}{named}{blue}
\definecolor{colKAK}{named}{magenta}
\definecolor{colMAB}{named}{Turquoise}
\definecolor{colMAX}{named}{Orange}

\newif\ifshowresolved
\showresolvedtrue

\usepackage{enumitem}
\usepackage{tikz}
\usetikzlibrary{positioning, arrows.meta, calc, fit, backgrounds}
\usepackage{pgfplots}          
\pgfplotsset{compat=1.18}
\usepgfplotslibrary{groupplots}

\usepackage{stackengine}
\usepackage{scalerel}

\usepackage{tcolorbox}
\tcbuselibrary{skins,theorems}

\usepackage{algorithm}
\usepackage{algpseudocode}
\usepackage{etoolbox}

\newcommand*{\algrule}[1][\algorithmicindent]{%
  \hspace*{.5em}\vrule\vrule
  width 0pt height \baselineskip depth .25\baselineskip
  \hspace*{\dimexpr#1-.5em}}
\makeatletter
\newcount\ALG@printindent@tempcnta
\def\ALG@printindent{%
  \ifnum \theALG@nested>0%
  \ifx\ALG@text\ALG@x@notext%
  \else
  \unskip
  \ALG@printindent@tempcnta=1
  \loop
    \algrule[\csname ALG@ind@\the\ALG@printindent@tempcnta\endcsname]%
    \advance \ALG@printindent@tempcnta 1
  \ifnum \ALG@printindent@tempcnta<\numexpr\theALG@nested+1\relax
  \repeat
  \fi
  \fi}%
\patchcmd{\ALG@doentity}{\noindent\hskip\ALG@tlm}{\ALG@printindent}{}%
  {\errmessage{failed to patch ALG@doentity}}
\makeatother
\AtBeginEnvironment{algorithmic}{\lineskip0pt}
\newcommand\NoNumberState[1]{{\renewcommand{\alglinenumber}[1]{}\State #1}}

\theoremstyle{definition} 
\newtheorem{rmk}{Remark}

\usetikzlibrary{external}
\usetikzlibrary{decorations.pathreplacing}
\usetikzlibrary{arrows.meta}
\showresolvedfalse

\hypersetup{
pdftitle={Preprint_VijayakumaranEtAl_2026},
pdfsubject={},
pdfauthor={Vijayakumaran},
pdfkeywords={constitutive modeling, thermodynamic consistency, neural networks, plasticity, spectral representation, eigenvalue degeneracy},
unicode=false,
}

\usepackage[font=small]{caption} 

\begin{document}

\maketitle

\begin{abstract}
Neural network material models can represent yield surfaces and hardening laws with far more flexibility than classical closed-form plasticity formulations, but training such a model end-to-end through an implicit return map introduces non-trivial challenges. This work presents a thermodynamically consistent neural constitutive model for rate-independent, isotropic elastoplasticity that jointly learns a flexible yield function together with its evolution laws. For non-associative flow, the yield function extends the convex flow potential by a jointly trained additive term, which renders the model thermodynamically consistent by construction. Isotropy and convexity are enforced simultaneously across all learnable components through a permutation-invariant, spectral (eigenvalue-based) representation. Training proceeds through a backward-Euler return map, with gradients propagated by the implicit function theorem together with an eigenvalue-degeneracy-aware scheme for safe differentiation. The robustness and generality of the approach are demonstrated by reproducing known constitutive models, including non-associative flow, learning material behavior from stress--strain data extracted from a full-field finite element simulation, and capturing unknown non-associative behavior from homogenized representative volume elements of a foam. Finally, the proposed neural plasticity model is evaluated within finite element simulations to solve boundary-value-problem under plane stress.
\end{abstract}

	
\section{Introduction}
\label{sec:intro}

Computational simulation of plastic materials is central to the design and analysis of a vast class of engineering systems subject to permanent, irreversible deformation history, ranging from metal forming and additive manufacturing to geo-mechanical engineering and structural analysis. 
At the core of these simulations is the constitutive model, which encodes the plastic material's response to the deformation history, and governs how it evolves under loading.
Conventionally, these constitutive models, be it micromechanically motivated or phenomenological, are formulated within rigorous physics-based frameworks that ensure thermodynamic consistency \citep{coleman1959ont,coleman1967ther} together with other important principles of constitutive modeling, e.g., objectivity \citep{noll1958ama}, also sometimes referred to as frame indifference \citep{coleman1967ther} along with the observed material symmetries.

These constitutive models address the three key ingredients of plasticity, namely an elastic storage mechanism, a yield criterion that delineates the onset of plasticity, and an evolution mechanism that describes how plasticity progresses upon yielding. Their formulation is typically parsimonious, relying on hand-crafted functional forms with a small number of parameters that are calibrated to capture specific material behaviors, such as yielding, hardening, and flow characteristics. The resulting models are well-understood, admit efficient return-mapping algorithms \citep{simo1985aun,ortiz1986ana,miehe1998cons,miehe1998form}, and satisfy thermodynamic consistency by construction. Yet this parsimony also limits their expressive power, and extending them to capture complex yield or hardening behaviors can require expert re-derivation and re-implementation.

\subsection{Overview on constitutive modeling with neural networks}

The emergence of machine learning as a tool for constitutive modeling offers a route to overcoming this limitation in model expressivity. Neural networks, with their universal approximation properties \citep{cybenko1989appr,hornik1991appr}, can represent a wide range of complex, nonlinear functions. The earliest applications date back to the pioneering work of \citet{ghaboussi1991know} in the early 1990s, who used feedforward neural networks (FNNs) to learn constitutive relations directly from data. Widespread adoption, however, gained momentum only more than two decades later, as part of a broader shift toward data-driven approaches in computational mechanics, e.g., \citep{bessa2017fram,liu2016self}. This shift was driven by advances in software ecosystems, automatic differentiation, and high-performance computing. Recurrent neural networks (RNNs), which carry an internal state that encodes the loading history, further enabled the modeling of path-dependent material behavior \citep{mozaffar2019deep,bonatti2021one,ghavamian2019acce}. However, both FNN and RNN approaches treat the constitutive response as a data-driven phenomenological input-output map, with no structural mechanism ensuring thermodynamic admissibility.

Initial attempts to incorporate thermodynamics into constitutive models did so by embedding the constraints into the training objective, adding penalty or regularization terms that reward non-negative dissipation on the sampled data.
The thermodynamics-based artificial neural networks (TANNs) of \citet{masi2021ther,masi2022mult} embed a thermodynamic potential explicitly but penalize negative dissipation rates during training and do not guarantee this by construction. The hybrid models proposed in \citep{settgast2020ahy,malik2022ahy} model the yield function and the evolution of internal state variables with neural networks.
\citet{vlassis2021sobo,vlassis2021comp} incorporate Sobolev-type training \citep{czarnecki2017sobo} and component-based level-set parametrization. \citet{weber2023phys} augment FNNs with physics-based training enhancements for rate-independent plasticity. \citet{he2022ther} infer internal state variables from the hidden state of a recurrent network and enforce the dissipation inequality through a penalty term in the loss. In each of these cases, thermodynamic consistency is either not considered explicitly or is incorporated only as a training objective, rather than being guaranteed by the model formulation itself. Thus, it is satisfied only over the sampled loading states without offering any guarantee at unseen loading conditions.

 A more robust way to ensure thermodynamic consistency is to build it into the model itself, using a classical constitutive framework as a foundation. In the case of hyperelasticity, this amounts to the existence of a free energy function, together with objectivity and appropriate material symmetries, cf. \citep{linka2021cons,fuhg2022lear,kalina2023fete,thakolkaran2022nneu}. Another important aspect in neural elasticity is the construction of polyconvex elastic potentials, building on the seminal contributions of \citet{ball1976conv,ball1977rema}. A maturing line of work uses input-convex neural networks (ICNNs) \citep{amos2016inpu} and closely related constraining mechanisms to guarantee polyconvexity \citep{klein2022poly,linden2023neur,tac2024benc} among many others. For the special case of isotropy, highly flexible polyconvex models based on ICNNs and principal stretches \citep{vijayakumaran2025cons,tepole2025poly} or signed singular values \citep{geuken2025ano, geuken2026mode} have been proposed.

Extending this approach to plasticity is more challenging. The model must simultaneously learn the yield surface that delineates the limits of the elastic regime, enforce thermodynamic consistency within the elastic regime, handle the evolution of internal state variables during plastic deformation, allow for non-associative flow when the material calls for it, and enforce material symmetry across every learnable component.
Of these, thermodynamic consistency within the elastic regime is the best-understood requirement, thanks to the maturity of (hyper)elastic modeling, while internal variable evolution remains the most architecturally demanding.

The evolution of internal variables is the architectural challenge shared by plasticity and viscoelasticity. Both classes require evolution equations that guarantee thermodynamic consistency, i.e., non-negative dissipation rates; these are nonlinear ordinary differential equations (ODEs) in time. A time integration of the ODEs is then needed, commonly implemented with forward (explicit Euler), backward (implicit/backward Euler), or more sophisticated schemes \citep{simo1998comp}.  Backward-Euler integration avoids the step-size stability limits of explicit schemes, allowing for larger time steps without solution divergence, but requires solving a nonlinear system at each step (typically via Newton--Raphson).
Differentiating through that solver during training requires either unrolling the Newton iterations or applying the implicit function theorem to obtain gradients, both of which add non-trivial complexity to the training procedure. Forward-Euler schemes are simpler to differentiate through and are therefore preferred in several neural inelasticity models \citep{holthusen2024theo,holthusen2026age} despite their conditional stability.

In the viscoelastic setting, where dissipation is governed by smooth rate-dependent potentials, the classical generalized standard materials (GSM) framework by \citet{halphen1975sur} built upon the pioneering works of \citet{coleman1963the,biot1965mech} provides the neural models their natural structure. In such models, the free energy and a convex, normalized, and stationary dual dissipation potential are each parametrized as ICNNs.\footnote{An equivalent formulation based on the free energy and the dissipation potential is also possible.} The thermodynamic forces follow from differentiating the free energy with respect to the state variables, and the evolution equations follow from differentiating the dual dissipation potential with respect to these forces. For time integration, several of these neural GSM models use backward Euler for its favorable stability properties \citep{flaschel2025conv,kalina2026aph,friedrichs2026prec}, but training through the resulting nonlinear solve has been reported to have numerical difficulties \citep{kalina2026aph, holthusen2026aco}, and a similar iterative solve in an earlier model was reported unstable and abandoned in favor of a neural ODE reformulation \citep{tac2023data}. Some works avoid the solve entirely by inferring the internal variable through an auxiliary network \citep{rosenkranz2024visc,asad2023ame,holthusen2026aco}, though \citet{friedrichs2026prec} report no significant improvement from this alternative.

Plasticity introduces an added challenge as it requires a yield criterion that defines an elastic domain, within which the internal state does not evolve. In a rate-independent setting, this introduces a sharp elastic-to-plastic transition governed by complementarity conditions rather than a smooth rate law. The return-mapping algorithm, an implicit elastic-predictor plastic-corrector scheme, satisfies this condition exactly at the end of each step and is unconditionally stable, though explicit schemes are used too \citep{simo1998comp}. Initial works on neural plasticity modeling focussed on learning the yield surface from yield-onset data \citep{fuhg2022elas} directly instead of from stress-strain datasets. Although several subsequent works formulate thermodynamically consistent neural models for learning plasticity \citep{fuhg2023modu,meyer2023ther,jadoon2025auto}, these operated in simpler settings, such as learning the hardening evolution laws while keeping the yield surface to a pre-defined von Mises form and demonstrating only on uniaxial loading paths, or by applying yield surface learning and the evolution laws in two stages \citep{fuhg2023modu}.

More recent works by \citet{boes2026acco} and \citet{velden2026ano} extend an inelastic constitutive artificial neural network framework \citep{holthusen2024theo} to plasticity, in rate-independent and rate-dependent settings respectively, proposing simultaneous learning of yield function and internal variables evolution laws. Both frameworks also define the yield function and flow potential as formally independent, admitting non-associative flow in principle, but each restricts its demonstrated examples to an associative evolution law, and both works report training difficulties and a high sensitivity to initialization or a need for pre-training. Interestingly, they do a mixed time integration in that the internal variables are evolved explicitly, while the plastic multiplier is solved implicitly through a Newton--Raphson scheme. In contrast, \citet{bleyer2025lear} adopts a different approach, formulating the elasto-plastic update itself as a convex optimization problem via implicit layers, but restricts the formulation to perfect plasticity in this work.

A cross-cutting architectural requirement shared by all learnable components is the enforcement of material symmetry, most commonly isotropy, as an inductive bias rather than a property to be learned from data. For an isotropic material, the yield function, the flow potential, and the free energy must all be isotropic functions of their symmetric tensor arguments. This is commonly achieved by expressing the potentials in terms of isotropic invariants \citep{boehler1977oni,riemer2026cons}. Convexity in the tensor argument is then ensured if the neural network is convex in the invariants and, in addition, non-decreasing in those invariants that are convex but nonlinear functions of the tensor, whereas no such restriction applies to invariants that are linear in the tensor \citep{klein2022poly,kalina2026aph}. Since these conditions are sufficient but not necessary for convexity, they restrict the representable functions to a subset of all convex isotropic potentials, thereby reducing expressivity. Alternatively, the potentials can be expressed directly in terms of a tensor's eigenvalues, requiring the function to be permutation-invariant in them so that isotropy holds by construction. A classical result of \citet{davis1957all} shows that the resulting isotropic function is convex in the tensor if and only if the underlying permutation-invariant function is convex in the eigenvalues, with no additional monotonicity requirement. For a learnable model, a single permutation-invariant, convex network of the eigenvalues therefore enforces isotropy and convexity jointly while representing the full class of convex isotropic potentials, circumventing the representational restriction imposed by the invariant-based construction. Prior work has enforced this permutation invariance in different ways: \citet{vijayakumaran2025cons} use a symmetric weight-sharing scheme over the principal stretches and their pairwise products for hyperelasticity; \citet{geuken2025ano} enforce permutation invariance by symmetrizing the function with respect to the action of the permutation group; \citet{jadoon2026ther} instead build on the Deep Sets framework \citep{zaheer2017deep,kimura2024onp}, composing an elementwise-convex network with a permutation-invariant pooling operation, at the cost of cross-terms between eigenvalues that the pooling structure cannot represent. \citet{jadoon2026ther} fit the yield function directly to calibration data and do not train through the return map, so their approach never exposes the interaction between an eigenvalue-based representation and the return map's Newton--Raphson iteration. Resolving the challenges therein is central to jointly learning the yield function and its evolution law directly from stress--strain data.

\subsection{Objectives and contributions of this work}
Motivated by these gaps, this work presents a unified, thermodynamically consistent neural constitutive modeling framework for rate-independent, isotropic elastoplasticity at infinitesimal strains. Built around non-associative plasticity with combined isotropic and kinematic hardening, the model recovers associative plasticity as a special case. Isotropy is enforced across all learnable components through permutation-invariant spectral representations of the governing tensors, while convexity is imposed by construction wherever it is required for thermodynamic consistency.

\begin{figure}[t]
    \centering
    \includegraphics[width=\textwidth]{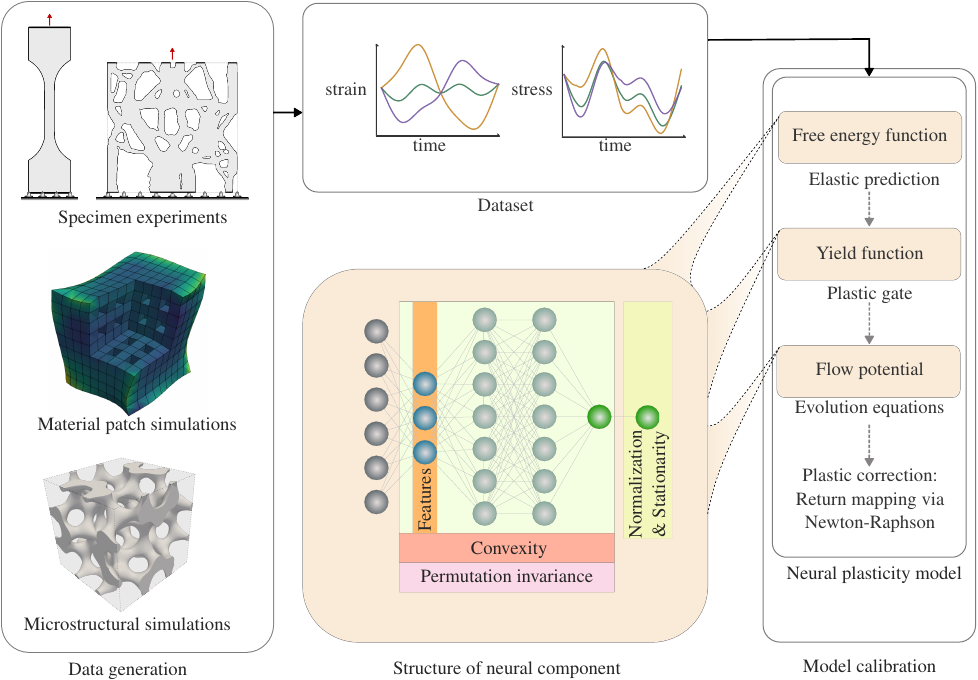}
    \caption{Overview of the proposed framework. Strain-stress data drawn from simulated specimen experiments \citep{yin2026desi}, material patch simulations \citep{ferreira2025auto}, or microstructural simulations \citep{abendroth2017ana,abendroth2024mode} calibrates a neural plasticity model in which convexity and permutation invariance are enforced component by component, as dictated by each network's own thermodynamic-consistency and isotropy requirements, within a differentiable return-mapping algorithm.}
    \label{fig:main_overview}
\end{figure}

The model's time integration is performed using a backward-Euler return map, propagating gradients through the Newton--Raphson scheme via the implicit function theorem rather than unrolling. A principled, degeneracy-safe solution is proposed for the eigenvalue-degeneracy singularity that can expose during training. The framework handles non-associative plasticity via an additive learnable network aimed at capturing the deviation of the yield function from the flow potential, while preserving thermodynamic consistency by construction. Figure~\ref{fig:main_overview} summarizes the overall pipeline.

The remainder of the paper is organized as follows.
Section~\ref{sec:plasticity basics} reviews the continuum framework of infinitesimal-strain plasticity and the return-mapping algorithm, establishing the notation used throughout.
Section~\ref{sec:neural_architecture} presents the neural network architecture and its coupling to the return map.
Section~\ref{sec:algorithmic} formulates training as an optimization problem constrained by the return mapping and addresses the differentiability challenge posed by eigenvalue degeneracy. Section~\ref{sec:numerical_results} evaluates the framework on known material laws, isotropy enforcement under plane stress, and non-associative flow in an open-cell foam material using a homogenized representative volume element (RVE) dataset. Section~\ref{sec:conclusion} concludes the paper and discusses directions for future work. The notation used within this work is introduced in Appendix~\ref{app:notation}.

\section{Thermodynamically consistent framework for infinitesimal-strain plasticity}\label{sec:plasticity basics}

This work focuses on rate-independent plasticity within the framework of infinitesimal strain theory. The formulation adheres to a thermodynamically consistent modeling framework that encompasses both non-associative and associative plasticity, accounting for mixed isotropic-kinematic hardening. In this section, we present the general governing equations of the model in Sections~\ref{subsec:free_energy}--\ref{subsec:associated} and outline the return mapping algorithm, which provides a robust numerical solution procedure in Section~\ref{subsec:return_mapping}.

\subsection{Helmholtz free energy function}\label{subsec:free_energy}

In line with classical plasticity theory, we assume an additive decomposition of the infinitesimal strain tensor into elastic and plastic parts, $\ueps = \ueps^\text{e} + \ueps^\text{p}$, with $\ueps,\ueps^\text{e},\ueps^\text{p}\in\Sym$.
Hardening is described by two strain-type hardening variables ($\ualpha\in\Sym$ and $\beta\in\SR_{\ge 0}$) associated with kinematic and isotropic hardening, respectively. The former tracks the (tensorial) translation of the yield surface, while the latter is a scalar measure of accumulated isotropic hardening.
For brevity, we collect the hardening variables in the tuple $\tuple H:=(\ualpha,\beta)$ and all strain-type internal variables in $\tuple I:=(\ueps^\text{p},\ualpha,\beta)=(\ueps^\text{p},\tuple H)$.
The free energy is then a mapping $\psi\colon\Sym\times\Sym\times\Sym\times\SR_{\ge 0}\to\SR, \;(\ueps,\tuple I)\mapsto\psi(\ueps,\tuple I)$.
We assume a split of the free energy into an elastic part and a plastic part collecting the hardening contributions, with the plastic part further splitting into kinematic and isotropic terms:
\begin{equation}
    \psi(\ueps,\tuple I) = \psi^\text{e}(\ueps-\ueps^\text{p}) + \psi^\text{p}(\tuple H) \; , \qquad
    \psi^\text{p}(\tuple H) = \psi^\text{kin}(\ualpha) + \psi^\text{iso}(\beta) \; .
    \label{eq:psi}
\end{equation}
We restrict attention to monotonically hardening behavior, meaning continued evolution of $\ualpha$ or $\beta$ in a fixed direction never diminishes the corresponding driving force. This is a modeling choice, not a thermodynamic requirement, since materials that soften under continued loading also exist. Section~\ref{sec:free_energy_architecture} shows how this hardening restriction is realized architecturally.
The Cauchy stress $\usig\in\Sym$ and the stress-type thermodynamic forces $\te A\in\Sym$, $B\in\SR_{\le 0}$ conjugate to the internal variables are defined by
\begin{align}
    \usig := \diffp{\psi}{\ueps} = -\diffp{\psi}{\ueps^\text{p}} = \diffp{\psi}{\ueps^\text{e}} \; , \quad
    \te A := -\diffp{\psi}{\ualpha} \; , \quad \text{and} \quad
    B := -\diffp{\psi}{\beta} \; .
    \label{eq:thermo_forces}
\end{align}
We summarize the thermodynamic forces in the tuple $\tuple F:=(\usig,\te A,B)$.\footnote{Since the yield function depends on $\usig$ and $\te A$ through the effective stress $\usig+\te A$ (Section~\ref{subsec:evolution_thermo}), $-\te A$ is the center of the yield surface in stress space and thus corresponds to the back-stress of classical kinematic hardening.} Since $\tuple F$ depends on $\ueps$ and $\tuple I$ through $\psi$, we write $\tuple F(\ueps,\tuple I)$ when we want to explicitly highlight this dependence.
To define the evolution of internal variables, we follow two classical frameworks, namely non-associative and associative plasticity theory, cf. the textbooks \citep{souzaneto2008comp,simo1998comp}. In the sections below, we discuss the more general non-associative case first, and then the associative case as a specialization.

\subsection{Evolution law and thermodynamic consistency}\label{subsec:evolution_thermo}
For describing the evolution of internal variables, we introduce a yield function $\varphi\colon \Sym \times \Sym \times \SR \to \SR,\; \tuple F \mapsto \varphi(\tuple F)$ and a flow potential $\chi\colon \Sym \times \Sym \times \SR \to \SR,\; \tuple F \mapsto \chi(\tuple F)$, sometimes called the plastic potential. While $\varphi$ determines \emph{whether} the material is yielding, $\chi$ determines \emph{in which direction} the internal variables evolve once it does. Formally, the yield function $\varphi$ defines the admissible domain
\begin{equation}
        \set E := \{
        \tuple F \in \Sym \times \Sym \times \SR \,
        | \, \varphi(\tuple F) \leq 0 \}
        \text{ and }
        \partial\set E := \{
        \tuple F \in \Sym \times \Sym \times \SR \,
        | \, \varphi(\tuple F) = 0 \} 
\end{equation}
in the thermodynamic-force space \citep{simo1998comp}.
For any driving force state $\tuple F \in \set{E} \setminus \partial\set{E} = \mathrm{int}(\set{E})$, i.e., in the elastic domain $\mathrm{int}(\set{E})$, the deformation behavior is purely elastic, and no plastic flow occurs. Plastic deformation can occur only when the driving force state reaches the boundary, i.e., $\tuple F\in \partial \set{E}$. Consistent with the roles described above, the flow potential $\chi$ governs the evolution of internal variables, given by
\begin{align}
    \dot{\ueps}^\text{p} = \lambda \diffp{\chi}{\usig} \; , \quad
    \dot{\ualpha} = \lambda \diffp{\chi}{\te A} \; , \quad
    \dot{\beta} = \lambda \diffp{\chi}{B} \; , \label{eq:evolution:non_associated}
\end{align}
subject to the Karush--Kuhn--Tucker (KKT) conditions
\begin{align}
    \lambda \ge 0 \; \wedge \; \varphi\le 0 \; \wedge \; \lambda \varphi = 0 \; ,
    \label{eq:KKT_conditions}
\end{align}
with $\lambda$ the plastic multiplier. By complementarity, $\lambda=0$ within the elastic domain $\mathrm{int}(\set{E})$, while $\lambda>0$ is possible only on the yield surface, i.e., when $\varphi=0$.

As shown in Appendix~\ref{app:thermo_consistency}, thermodynamic consistency is ensured if $\chi$ is convex in $\tuple F$ and satisfies
\begin{equation}
    \chi(\tuple F) \ge \chi(\te 0, \te 0, 0) \qquad \forall\, \tuple F \in \partial\set{E} \; .
\end{equation}
Since $\chi$ enters the evolution law only through its gradient, the value $\chi(\te 0,\te 0,0)$ can be chosen freely. We adopt the normalization $\chi(\te 0,\te 0,0)=0$, so that the condition becomes
\begin{equation}
    \chi(\tuple F) \ge 0 \qquad \forall\, \tuple F \in \partial\set{E} \; ,
    \label{eq:dissipation_condition_yield}
\end{equation}
i.e., $\chi$ is non-negative on the yield surface. Note that this condition only restricts $\chi$ on the boundary $\partial\set E$. Within the elastic domain $\mathrm{int}(\set E)$, the KKT conditions give $\lambda=0$, so the sign of $\chi$ there does not enter the dissipation rate. The same condition applies to both the non-associative and associative cases, as discussed in Sections~\ref{subsec:non-associated} and~\ref{subsec:associated}, respectively.

Phenomenologically, we choose the yield function and flow potential such that ideal plasticity, isotropic hardening, and kinematic hardening are handled through a modular construction. Let $\tilde\chi$ and $\tilde\varphi$ denote the stress-dependent parts of the flow potential and the yield function in an ideal-plasticity setting. The kinematic hardening force $\te A$ enters through the effective stress $\usig^\text{eff}(\usig,\te A):=\usig+\te A$. The isotropic hardening force $B$ enters additively, such that the yield stress $\sigma_\text{y}(B):=\sigma_{\text{y}_0}-B$ increases from its initial value $\sigma_{\text{y}_0}$ as hardening progresses, where $B\le0$ under the monotonic hardening assumed in Section~\ref{subsec:free_energy}. The functions take the form
\begin{align}
    \chi(\tuple F) &:= \tilde\chi\big(\usig^\text{eff}(\usig,\te A)\big) + B \; , \label{eq:flow_potential} \\
    \varphi(\tuple F) &:= \tilde\varphi\big(\usig^\text{eff}(\usig,\te A)\big) - \sigma_\text{y}(B) \; . \label{eq:yield_function}
\end{align}
At the force-free state, $\usig=\te A=\te 0$ and $B=0$, so $\chi(\te 0,\te 0,0)=\tilde\chi(\te 0)=0$. Sections~\ref{subsec:non-associated} and~\ref{subsec:associated} specialize $\tilde\chi$ and $\tilde\varphi$ to the non-associative and associative cases, respectively, where the non-associative case additionally extends the flow potential of Eq.~\eqref{eq:flow_potential} by hardening terms.

\begin{rmk}
    \label{remark:isotropy}
    This work restricts attention to isotropic materials, meaning the constitutive response does not depend on the orientation of the material. The free energy, flow potential, and yield function introduced above must therefore all be isotropic tensor functions of their arguments. For example, the free energy must satisfy $\psi(\te Q \star \ueps, \te Q \star \tuple I) = \psi(\ueps, \tuple I) \; \forall \te Q \in \Orth, \ueps \in \Sym, \tuple I \in \Sym\times\Sym\times \SR_{\ge 0}$ \citep{riemer2026cons}. Therein, $\star$ denotes the group action on a tensor or the tensors in a tuple, e.g., $\te Q \star \ueps := \te Q \cdot \ueps \cdot \te Q^T$. The isotropy requirements for $\chi$ and $\varphi$ are analogous, and apply equally to the non-associative and associative cases below.
\end{rmk}

\subsection{Non-associative plasticity}\label{subsec:non-associated}

In the non-associative formulation, $\tilde\chi$ and $\tilde\varphi$ may differ. This means that the flow direction, described by $\chi$, does not have to be normal to the yield surface $\varphi=0$. Since the thermodynamic argument requires convexity of $\chi$ alone, $\varphi$ is not required to be convex.

We construct $\tilde\varphi$ by combining the flow potential's convex part $\tilde \chi$ with an additional term $\Delta$, i.e.,
\begin{equation}
\tilde{\varphi}(\usig^\text{eff}) = \tilde{\chi}(\usig^\text{eff}) + \Delta(\usig^\text{eff}) \; .
\label{eq:F_split}
\end{equation}
Apart from isotropy, $\Delta$ is subject only to an upper bound and is non-convex in general, so that $\tilde\varphi$, and hence $\varphi$, may be non-convex as well. Appendix~\ref{app:TC_non_associated} shows that this restriction satisfies Eq.~\eqref{eq:dissipation_condition_yield} if $\Delta(\usig^\text{eff})<\sigma_{\text{y}_0}$ for all $\usig^\text{eff}\in\Sym$.

To increase the expressivity of the hardening evolution under non-associativity, we extend the flow potential of Eq.~\eqref{eq:flow_potential} by additional terms for kinematic and isotropic hardening, i.e.,
\begin{align}
    \chi(\tuple F) := \tilde{\chi}(\usig^\text{eff}) + B + \chi^\text{kin}(\te A) + \chi^\text{iso}(B) \; ,
    \label{eq:flow_potential_extended}
\end{align}
where $\chi^\text{kin}$ and $\chi^\text{iso}$ are non-negative, convex functions of their respective inputs with $\chi^\text{kin}(\te 0)=\chi^\text{iso}(0)=0$. Thus, $\chi$ remains convex in $\tuple F$ and the normalization $\chi(\te 0,\te 0,0)=0$ is preserved. These terms modify the evolution of $\ualpha$ and $\beta$, respectively. The associative case of Section~\ref{subsec:associated} is recovered exactly when $\Delta\equiv0$ and $\chi^\text{kin}\equiv\chi^\text{iso}\equiv0$.

A more general formulation could also account for mixed terms coupling $\usig$, $\te A$, and $B$, but this would require mixed isotropic invariants \citep{boehler1977oni} for a more complex architecture. Since the additive decomposition into individual contributions already proved sufficient for the present purpose, we adopt this simpler, more comprehensible structure instead.

\begin{rmk}[Alternative non-associative formulation]
\label{remark:non_associative_alternative}

A broader restriction on the flow potential $\chi$ than the thermodynamic condition of Eq.~\eqref{eq:dissipation_condition_yield} can make $\varphi$ and $\chi$ independent of each other. Requiring $\chi\ge0$ everywhere, instead of only on the yield surface, removes the need for the shared relationship between $\varphi$ and $\chi$ that we adopt in Eq.~\eqref{eq:F_split}. This condition, like Eq.~\eqref{eq:dissipation_condition_yield} itself, is sufficient but not necessary for $D\ge0$. However, for a flow potential $\chi$ that is convex and normalized ($\chi(\te 0,\te 0,0)=0$) to be non-negative everywhere, the force-free state must be a global minimum of $\chi$. This implies that $\chi$ must be stationary at that point, i.e., its gradient must vanish at the force-free state. Enforcing this architecturally constrains the type of pressure-sensitive flow $\chi$ can represent, as discussed in Section~\ref{sec:flow_potential}. This representational limitation and the slower convergence observed when $\chi$ and $\varphi$ are parametrized by independent networks led us to adopt the additive formulation of Eq.~\eqref{eq:F_split} throughout this work.
\end{rmk}

\subsection{Associative plasticity}\label{subsec:associated}
Associative plasticity is the special case in which the flow potential and the yield function are identified up to an additive constant, so that their gradients $\partial \chi /\partial {\tuple F}\equiv \partial \varphi /\partial {\tuple F}$ coincide, and the flow direction is normal to the yield surface $\varphi=0$ for every internal variable. Here, we set $\Delta\equiv0$ in Eq.~\eqref{eq:F_split}, i.e., $\tilde\varphi(\usig^\text{eff}) = \tilde\chi(\usig^\text{eff})$, and use the flow potential of Eq.~\eqref{eq:flow_potential} without the additional hardening terms of Eq.~\eqref{eq:flow_potential_extended}. Inserting this into Eqs.~\eqref{eq:flow_potential} and~\eqref{eq:yield_function} gives
\begin{equation}
    \chi(\tuple F) = \tilde\chi(\usig^\text{eff}) + B = \tilde\varphi(\usig^\text{eff}) - \sigma_\text{y}(B) + \sigma_{\text{y}_0} = \varphi(\tuple F) + \sigma_{\text{y}_0} \; ,
\end{equation}
so that $\chi$ and $\varphi$ differ only by the constant $\sigma_{\text{y}_0}$, which gives the gradient identification stated above. In particular, $\tilde\varphi(\te 0)=\tilde\chi(\te 0)=0$ by the normalization adopted in Section~\ref{subsec:evolution_thermo}. The evolution of internal variables is
\begin{align}
    \dot{\ueps}^\text{p} = \lambda \diffp{\varphi}{\usig} \; , \quad
    \dot{\ualpha} = \lambda \diffp{\varphi}{\te A} \; , \quad
    \dot{\beta} = \lambda \diffp{\varphi}{B} \; ,
\end{align}
subject to the same KKT conditions given in Eq.~\eqref{eq:KKT_conditions}.

The thermodynamic-consistency condition of Eq.~\eqref{eq:dissipation_condition_yield} applies unchanged. Since $\varphi=0$ for $\tuple F \in \partial\set E$, Eq.~\eqref{eq:dissipation_condition_yield} gives $\chi=\sigma_{\text{y}_0}$ there. As shown in Appendix~\ref{app:TC_associated}, this reduces the condition to requiring convexity of $\varphi$ in $\tuple F$ together with
\begin{equation}
    \sigma_{\text{y}_0} \ge 0 \; , \qquad \text{i.e.,} \qquad \varphi(\te 0, \te 0, 0) = -\sigma_{\text{y}_0} \le 0 \; ,
\end{equation}
which means the force-free state must lie in the admissible domain $\set E$.

\subsection{Return-mapping algorithm}\label{subsec:return_mapping}

The time interval $[0,T]$ is divided into sub-intervals $[{}^{(n-1)}t,\,{}^{n}t]$ with step size $\tau := {}^{n}t - {}^{(n-1)}t > 0$. Field variables at step $(n-1)$ are assumed to be known. A quantity without a left superscript is understood to be at step $n$, while a left superscript $(n-1)$ denotes one at the previous step. For instance, $\ueps$ denotes ${}^{n}\ueps$ and $\tuple I$ denotes ${}^{n}\tuple I$; this convention holds throughout the present subsection and matches the step index $n$ used in Section~\ref{subsec:optimisation_problem} and Figure~\ref{fig:training_loop}.
To numerically solve the evolution of internal variables, we use a predictor-corrector method \citep{simo1985aun,ortiz1986ana,simo1998comp,souzaneto2008comp}.


\begin{algorithm}[b!]
\caption{Local return-mapping algorithm for the neural plasticity model.}
\label{ALG_RMA}
\begin{algorithmic}[1]
\Require $\ueps$,\enspace ${}^{(n-1)}\tuple I$
\State $\tuple I \gets {}^{(n-1)}\tuple I$;\enspace
  $\gamma \gets 0$
  \Comment{initialize}
\State Compute the driving forces and flow-potential gradient:
\NoNumberState{\hspace*{1em}
  $\tuple F \gets \tuple F(\ueps,\tuple I)$;\enspace
  $\tuple V$ $\gets \diffp{\chi}{\tuple F}\Big|_{\tuple F}$}
\If{$\varphi(\tuple F) \le 0$}
  \Comment{elastic step}
  \State \Return $\tuple F$,\enspace $\tuple I$
\Else
  \Comment{plastic step}
  \State \textbf{compute} residual:
  \NoNumberState{\hspace*{1em}
    $\tuple R \gets
    \begin{bmatrix}
      \tuple I - {}^{(n-1)}\tuple I - \gamma\,\tuple V \\[2pt]
      \varphi(\tuple F)
    \end{bmatrix}$}
  \While{$\|\tuple R\| \geq \text{tol}_\mathrm{N}$}
    \Comment{Newton iteration}
    \State \textbf{compute} Newton step:
    \NoNumberState{\hspace*{1em} solve $\tuple R + \tuple J\,\Delta \tuple X = 0$\enspace for
      $\Delta \tuple X = [\Delta\tuple I,\,\Delta\gamma]^\top$}
    \NoNumberState{\hspace*{1em} with Jacobian $\tuple J = \partial\tuple R/\partial \tuple X$ via automatic differentiation}
    \State \textbf{update} variables:
    \NoNumberState{\hspace*{1em}
      $\tuple I \gets \tuple I + \Delta\tuple I$;\quad
      $\gamma \gets \gamma + \Delta\gamma$}
    \State \textbf{re-evaluate} driving forces, gradient, and residual:
    \NoNumberState{\hspace*{1em}
      $\tuple F \gets \tuple F(\ueps,\tuple I)$;\enspace
      $\tuple V$ $\gets \diffp{\chi}{\tuple F}\Big|_{\tuple F}$}
    \NoNumberState{\hspace*{1em}
      $\tuple R \gets
      \begin{bmatrix}
        \tuple I - {}^{(n-1)}\tuple I - \gamma\,\tuple V \\[2pt]
        \varphi(\tuple F)
      \end{bmatrix}$}
  \EndWhile
  \State \Return $\tuple F$,\enspace $\tuple I$
\EndIf
\Ensure $\tuple F = \tuple F(\ueps,\tuple I)$ and $\tuple I$
\end{algorithmic}
\end{algorithm}

In the predictor step, the internal variables are held constant, and the yield condition $\varphi(\tuple F^\text{trial})\le 0$ is checked, where $\tuple F^\text{trial}:=\tuple F(\ueps,{}^{(n-1)}\tuple I)$ denotes the thermodynamic forces of Eq.~\eqref{eq:thermo_forces} evaluated at the current strain and the internal variables of the previous step. If the yield condition is satisfied, the step is elastic and the internal variables remain unchanged. Otherwise, the corrector step applies a return-mapping algorithm to update the internal variables.

\begin{figure}[tb]
    \centering
    \resizebox{\linewidth}{!}{\includegraphics{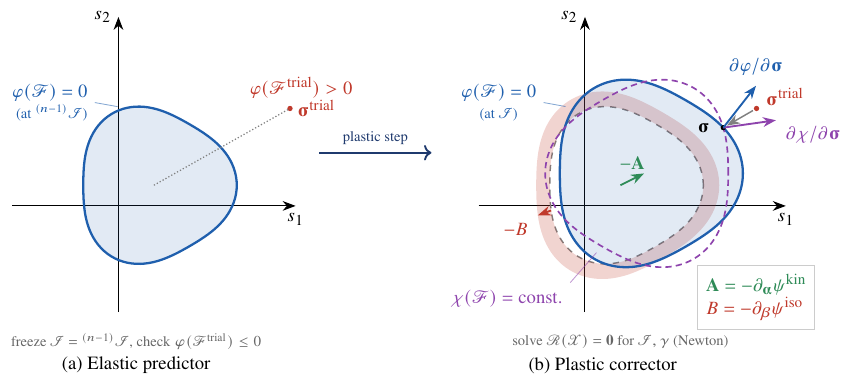}}
    \caption{Return mapping, illustrated in a two-dimensional section of stress space with coordinates $s_1$ and $s_2$. In (a), the elastic predictor freezes $\tuple I$ at ${}^{(n-1)}\tuple I$ and checks the trial forces $\tuple F^{\mathrm{trial}}=\tuple F(\ueps,{}^{(n-1)}\tuple I)$ against the yield surface. If $\varphi(\tuple F^{\mathrm{trial}})\le0$, the step is elastic and $\tuple I={}^{(n-1)}\tuple I$. Otherwise, the plastic corrector in (b) applies. Kinematic hardening shifts the center of the yield surface by $-\te A$, and isotropic hardening expands it by $-B\ge0$, moving it from its state at ${}^{(n-1)}\tuple I$ (dashed gray) to its converged state at $\tuple I$ (solid blue). Simultaneously, the trial stress $\usig^{\mathrm{trial}}$ is returned to the converged stress $\usig$ on the updated yield surface. The correction is driven by the flow direction $\partial\chi/\partial\usig$, evaluated at the converged state and normal to the level set $\chi(\tuple F)=\mathrm{const.}$ through that state (dashed purple). In the non-associative case, this direction differs from the yield surface normal $\partial\varphi/\partial\usig$, a consequence of $\partial\chi/\partial\tuple F\neq\partial\varphi/\partial\tuple F$.}
    \label{fig:return_mapping}
\end{figure}

For the corrector step, we apply an implicit backward Euler scheme to the evolution laws given in Eq.~\eqref{eq:evolution:non_associated}, together with the consistency condition $\varphi=0$ implied by the KKT conditions~\eqref{eq:KKT_conditions} for $\lambda>0$. We thus get the discrete system
\begin{align}
    \tuple I - {}^{(n-1)}\tuple I &= \gamma \tuple V \quad \text{with} \quad
    \tuple V := \diffp{\chi}{\tuple F}\Big|_{\tuple F(\ueps, \tuple I)} \quad \text{and} \quad
    \gamma := \tau \lambda \; ,\label{eq:return-map1}\\
    \varphi(\tuple F(\ueps, \tuple I)) &= 0 \; . \label{eq:return-map2}
\end{align}
Here, $\tuple V$ denotes the flow direction, i.e., the gradient of the flow potential with respect to the thermodynamic forces $\tuple F$ evaluated at the converged state. Since $\partial\chi/\partial\tuple F=\partial\varphi/\partial\tuple F$ in the associative case (Section~\ref{subsec:associated}), the same scheme applies to both formulations. Stacking Eqs.~\eqref{eq:return-map1} and~\eqref{eq:return-map2} gives the residual $\tuple R \in \Sym \times \Sym \times \SR \times \SR$, required to vanish up to a given tolerance. With the unknown $\tuple X := (\tuple I, \gamma)$ and the Jacobian $\tuple J := \partial\tuple R/\partial\tuple X$, Newton's method updates $\tuple X$ via $\tuple R + \tuple J\,\Delta \tuple X = 0$. 
The local return-mapping scheme summarized in Algorithm~\ref{ALG_RMA} is an adaptation of the algorithm presented by \citet{miehe1998cons,miehe1998form}. Figure~\ref{fig:return_mapping} illustrates the predictor and corrector steps geometrically in stress space.

\begin{rmk}[Plane stress]
\label{remark:plane_stress}
The algorithm above treats $\ueps$, $\tuple I$, and $\tuple F$ as whole tensor- and tuple-valued quantities. This suffices whenever the entire strain tensor $\ueps$ is prescribed as input, as is the case in three dimensions. Under the plane-stress condition $\sigma_{33}=0$, only the in-plane strain is prescribed, so the algorithm must instead be written component-wise, with each remaining component playing a different role, either as a new unknown, an eliminated quantity, or a retained unknown. The out-of-plane normal strain $\varepsilon_{33}$ becomes a new unknown, solved jointly with $\tuple I$ and $\gamma$ subject to the added condition $\sigma_{33}=0$ in $\tuple R$. The out-of-plane shear strains are instead eliminated, since isotropy of $\chi$ and the reflection symmetry of the plane make the out-of-plane shear components of the flow direction vanish whenever the corresponding components of $\usig$ and $\te A$ are zero. Since these components start at zero, they remain identically zero throughout the loading history. The plastic thickness strain $\varepsilon^\text{p}_{33}$, already part of $\tuple I$, is retained as an unknown, since it is generally non-zero.
\end{rmk}

\section{Neural architecture for thermodynamically consistent plasticity}\label{sec:neural_architecture}

The continuum model developed in Section~\ref{sec:plasticity basics} introduces additive decompositions and constraints on the sub-potentials, leaving the functional forms of the free energy $\psi$, the flow potential $\chi$, and the yield function $\varphi$ unspecified. In contrast, the additive strain decomposition, the evolution laws of Eq.~\eqref{eq:evolution:non_associated}, the KKT conditions of Eq.~\eqref{eq:KKT_conditions}, and the return-mapping scheme of Algorithm~\ref{ALG_RMA} are already fixed.
This section describes how the introduced sub-functions are parameterized using neural networks in such a way that the structural requirements from Section~\ref{sec:plasticity basics} (namely, convexity, isotropy, normalization, and stationarity at the zero states) are enforced at the architectural level, thereby ensuring thermodynamic consistency. The presentation is given for the non-associative case. The associative case is recovered by setting $\Delta\equiv0$ and omitting $\chi^\text{kin}$ and $\chi^\text{iso}$, cf. Section~\ref{subsec:associated}.

\subsection{Design philosophy}
\label{sec:design_philosophy}

Before turning to the specification of the sub-functions, we describe the enforcement of convexity, isotropy, as well as normalization and stationarity at the zero states in a scalar-valued neural network.

\subsubsection{Convexity}
\label{sec:icnn}

Convexity of a learned scalar function is achieved using input-convex neural networks (ICNNs)~\citep{amos2016inpu}. As appropriate to the learned function, its arguments could be scalar quantities, the components of a symmetric tensor, or, when enforcing isotropy of a function of a symmetric tensor (cf. Section~\ref{sec:symmetry}), the eigenvalues of that tensor. A convex, non-decreasing activation is applied at every layer, here the softplus function $\mathcal{S\!P}(x) = \ln(1+e^x)$. The ICNN architecture achieves convexity in the inputs by constraining the weights of all but the first layer to be non-negative. The first layer is convex for any weights, and each further layer preserves convexity, since non-negative weighted sums of convex functions composed with a convex, non-decreasing activation remain convex~\citep{boyd2004conv}.

\begin{rmk}
    The original ICNN formulation by \citet{amos2016inpu} also includes skip connections from the input to every layer. These were not found to add substantial expressivity in the present setting and are therefore not used.
\end{rmk}

\subsubsection{Isotropy}
\label{sec:symmetry}

For enforcing isotropy on learned scalar functions of a symmetric second-order tensor argument $\te T\in\Sym$, we define a spectral representation in terms of the eigenvalues $(T_1, T_2, T_3)\in\SR^3$ of $\te T$. The isotropy requirements of Remark~\ref{remark:isotropy} then reduce, in each case, to invariance of this function under the permutation group $\tuple S_3$ acting on $(T_1, T_2, T_3)$. This symmetry is enforced by permutation averaging (Janossy pooling)~\citep{murphy2019jano}. A scalar network $g^{\mathrm{NN}}\colon \SR^3 \to \SR$ without built-in symmetry is evaluated on each of the six permuted eigenvalue triples, and the results are averaged, i.e.,
\begin{equation}
\bar{g}^{\mathrm{NN}}(T_1, T_2, T_3) := \frac{1}{|\tuple S_3|}\sum_{\pi \in \tuple S_3} g^{\mathrm{NN}}\bigl(T_{\pi(1)}, T_{\pi(2)}, T_{\pi(3)}\bigr) \; , \qquad |\tuple S_3| = 6 \; .
\label{eq:permutation_averaging}
\end{equation}

Choosing $g^{\mathrm{NN}}$ in Eq.~\eqref{eq:permutation_averaging} as an ICNN according to Section~\ref{sec:icnn} makes $\bar{g}^{\mathrm{NN}}$ convex in the eigenvalues as well as permutation-invariant. However, the convexity condition required by the thermodynamic argument of Appendix~\ref{app:thermo_consistency} is convexity in the tensor argument $\te T$ itself.
By Davis's theorem \citep{davis1957all,lewis1996conv}, a function of the eigenvalues induces a convex function of $\te T$ if and only if it is both convex and permutation-invariant. Since $\bar g^{\mathrm{NN}}$ has both properties, the spectral function $\bar g^\text{NN}\colon\Sym\to\SR,\;\te T\mapsto\bar g^\text{NN}\bigl(T_1(\te T),T_2(\te T),T_3(\te T)\bigr)$ is convex and isotropic in $\te T$. Moreover, the permutation-invariant spectral function is differentiable in $\te T$ whenever $\bar g^\text{NN}$ is differentiable, including at repeated eigenvalues \citep{lewis1996deri,miehe1993comp}.

\begin{rmk}
    Janossy pooling can optionally be generalized by making $\bar{g}^{\mathrm{NN}}$ vector-valued and mapping the averaged, isotropic feature vector through a second, outer network for extra expressivity. This was not found necessary here and is thus not used in the present work.
\end{rmk}

\begin{rmk}[Triclinic alternative]
\label{remark:triclinic}
When isotropy enforcement is not desired, the function can instead be represented directly via the 6 independent components of the symmetric $\te T$ in Mandel notation with respect to a fixed basis $\{\ve e_k\}_{k=1}^3$, i.e.,
\begin{equation}
g^\text{NN}\colon \te T \mapsto g^\text{NN}\bigl(T_{11}(\te T),T_{22}(\te T),T_{33}(\te T),\sqrt{2}\,T_{23}(\te T),\sqrt{2}\,T_{13}(\te T),\sqrt{2}\,T_{12}(\te T)\bigr) \; .
\label{eq:triclinic_component}
\end{equation}
These 6 components form an invariant set that is a functional basis for the triclinic case \citep{riemer2026cons}. Since this is a linear reparametrization of $\te T$, convexity of $g^\text{NN}$ in these 6 components directly implies convexity in $\te T$ itself, but the resulting function is not isotropic in general.
\end{rmk}
\subsubsection{Normalization and stationarity at zero states}
\label{sec:zero_states}

Since the hardening part of the free energy, the flow potential, and the yield function are represented by neural networks, we must ensure that the learned functions behave appropriately at their respective zero states, i.e., the reference state for the free energy and the force-free state for the flow potential and the yield function. Two types of corrections are used: a \emph{value correction} (normalization), enforcing $g(\te 0)=0$, and a \emph{gradient correction} (stationarity), additionally enforcing $\partial g/\partial\te{T}\big|_{\te{T}=\te{0}}=\te{0}$. For a tensor argument $\te T$ with eigenvalues $(T_1,T_2,T_3)$, the corrected function reads 
\begin{equation}
    g(\te T) := \bar g^\text{NN}(T_1(\te T),T_2(\te T),T_3(\te T))  - \bar g^\text{NN}(0,0,0)  -\sum_{i=1}^{3}\diffp{\bar{g}^{\mathrm{NN}}}{T_i}\bigg|_{\te T = \te 0} T_i(\te T) \; ,
    \label{eq:zero_state_corrections}
\end{equation}
and analogously, without permutation averaging, for scalar arguments or when $g^\text{NN}$ acts directly on the components of $\te T$ (cf. Remark~\ref{remark:triclinic}). Since $\bar g^\text{NN}$ is permutation-invariant, all partial derivatives $\partial\bar g^\text{NN}/\partial T_i$ coincide at the origin, so that the gradient correction is proportional to $T_1+T_2+T_3=\tr\te T$. Both corrections are thus affine and isotropic, and they preserve the convexity and isotropy of $\bar g^\text{NN}$. Which correction is applied to which component is specified in Sections~\ref{sec:free_energy_architecture}--\ref{sec:yield_function}.

\begin{rmk}
    For a convex function, the two corrections together place the global minimum at the origin with value zero, so the function is non-negative everywhere. Normalization alone does not ensure this. 
\end{rmk}

\subsection{Learnable components and their requirements}

This section restates the learnable components and summarizes the structural requirements established in Section~\ref{sec:plasticity basics}, which the following sections (Sections~\ref{sec:free_energy_architecture}--\ref{sec:yield_function}) realize architecturally.
As the focus of the present work is on the modeling and identification of plasticity, the elastic response is assumed to be known and fixed. As a result, only the plastic part $\psi^\text{p}$ of the free energy, the flow potential $\chi$, and the yield function $\varphi$ are learned. 
For reference in what follows, we restate these learnable components 
\begin{align}
\psi^\text{p}(\tuple H) &= \psi^\text{kin}(\ualpha) + \psi^\text{iso}(\beta) \; ,\\
\chi(\tuple F) &= \tilde{\chi}(\usig^\text{eff}) + B + \chi^\text{kin}(\te A) + \chi^\text{iso}(B) \; , \label{eq:G_full} \\
\varphi(\tuple F) &= \tilde{\varphi}(\usig^\text{eff}) + B - \sigma_{\text{y}_0} \label{eq:F_full} \\
&= \tilde{\chi}(\usig^\text{eff}) + \Delta(\usig^\text{eff}) + B - \sigma_{\text{y}_0} \; , \label{eq:F_full_delta}
\end{align}
of the free energy from Eq.~\eqref{eq:psi}, as well as the flow potential and the yield function from Eqs.~\labelcref{eq:yield_function,eq:F_split,eq:flow_potential_extended}. Therein, $\sigma_{\text{y}_0} > 0$ is also a trainable initial yield threshold.
Table~\ref{tab:design_summary} collects the structural requirements on the learnable components introduced in this section and their physical or thermodynamic origins.

\begin{table}[t]
\centering
\small
\caption{Architectural requirements on the learnable components and their physical or thermodynamic origins.}
\label{tab:design_summary}
\begin{tabular}{p{2.8cm}p{6.0cm}p{5.0cm}}
\toprule
Component & Constraint & Origin \\
\midrule
$\psi^\text{iso}(\beta)$
    & Convex; $\psi^\text{iso}(0)=0$, stationary at $0$; hence non-negative and $B\le0$ for $\beta\ge0$
    & Monotonic isotropic hardening; $B$ vanishes in the reference state (Section~\ref{subsec:free_energy}) \\[4pt]
$\psi^\text{kin}(\ualpha)$
    & Convex; $\psi^\text{kin}(\te 0)=0$, stationary at $\te 0$; hence non-negative
    & Monotonic kinematic hardening in every direction; $\te A$ vanishes in the reference state (Section~\ref{subsec:free_energy}) \\[4pt]
$\tilde\chi(\usig^\text{eff})$
    & Convex; $\tilde\chi(\te 0)=0$; gradient at $\te 0$ left free
    & Convexity of $\chi$ and normalization $\chi(\te 0,\te 0,0)=0$ (Section~\ref{subsec:evolution_thermo}); pressure sensitivity \\[4pt]
$\chi^\text{kin}(\te A)$, $\chi^\text{iso}(B)$
    & Convex; $\chi^\text{kin}(\te 0)=\chi^\text{iso}(0)=0$, stationary at $\te 0$/$0$; hence non-negative; optional, off by default
    & Non-associative kinematic/isotropic hardening evolution; convexity of $\chi$ and $\chi\ge0$ on $\partial\set E$ (Section~\ref{subsec:non-associated}) \\[4pt]
$\Delta(\usig^\text{eff})$
    & No convexity requirement; $\Delta<\sigma_{\text{y}_0}$ for all $\usig^\text{eff}$, no lower bound; $\Delta(\te 0)=0$
    & Non-associative flow; $\chi>0$ on $\partial\set E$ (Appendix~\ref{app:TC_non_associated}) \\[4pt]
$\sigma_{\text{y}_0}$
    & Positive
    & Force-free state in the elastic domain $\mathrm{int}(\set E)$ \\[4pt]
$\tilde\chi,\,\Delta,\,\psi^\text{kin},\,\chi^\text{kin}$
    & Isotropic, via permutation averaging over eigenvalues
    & Material symmetry (Remark~\ref{remark:isotropy}) \\
\bottomrule
\end{tabular}
\end{table}

\begin{rmk}
    If the elastic response is unknown, it can either (a) be learned first from purely elastic data and fixed before training the plastic components, or (b) be identified jointly with the plastic components in an iterative elastic-plastic co-identification procedure. The present work focuses on option (a), but the architecture of the learned plastic components is the same for both options.
\end{rmk}

\subsection{Free energy function}
\label{sec:free_energy_architecture}

In this work, the elastic free energy $\psi^\text{e}$ of Eq.~\eqref{eq:psi} is taken as an isotropic quadratic function of $\ueps^\text{e}$, expressed in terms of the first invariant of $\ueps^\text{e}$ and the second invariant of its deviator, i.e.,
\begin{equation}
\psi^\text{e}(\ueps^\text{e}) = \tfrac{1}{2}\kappa\, \bigl(\tr[\ueps^\text{e}]\bigr)^2 + \mu\, \tr\bigl[(\dev[\ueps^\text{e}])^2\bigr] \; .
\label{eq:psi_elastic_linear}
\end{equation}
Here, $\kappa \in \SR_{>0}$ denotes the bulk modulus and $\mu \in \SR_{>0}$ the shear modulus. This form is strictly convex in $\ueps^\text{e}$, which ensures a positive-definite elastic stiffness.

Isotropic and kinematic hardening each contribute a free-energy term, $\psi^\text{iso}$ and $\psi^\text{kin}$, each built from its own ICNN (Section~\ref{sec:icnn}). The network $g^{\mathrm{NN}}_{\psi^\text{iso}}$ acts on the scalar $\beta$, while $\bar{g}^{\mathrm{NN}}_{\psi^\text{kin}}$ acts on the eigenvalues $(\alpha_1,\alpha_2,\alpha_3)\in\SR^3$ of $\ualpha$ and is permutation-invariant (Section~\ref{sec:symmetry}). Both receive the value and gradient corrections of Section~\ref{sec:zero_states}, i.e.,
\begin{align}
\psi^\text{iso}(\beta) &= g^{\mathrm{NN}}_{\psi^\text{iso}}(\beta) - g^{\mathrm{NN}}_{\psi^\text{iso}}(0) - \left.\frac{\mathrm{d}g^{\mathrm{NN}}_{\psi^\text{iso}}}{\mathrm{d}\beta}\right|_{\beta=0}\beta \; ,
\label{eq:psi_iso_integral} \\[4pt]
\psi^\text{kin}(\ualpha) &= \bar{g}^{\mathrm{NN}}_{\psi^\text{kin}}\bigl(\alpha_1(\ualpha),\alpha_2(\ualpha),\alpha_3(\ualpha)\bigr) - \bar{g}^{\mathrm{NN}}_{\psi^\text{kin}}(0,0,0) - \sum_{i=1}^{3} \left.\frac{\partial \bar{g}^{\mathrm{NN}}_{\psi^\text{kin}}}{\partial \alpha_i}\right|_{\ualpha=\te{0}}\alpha_i(\ualpha) \; ,
\label{eq:psi_kin_icnn}
\end{align}
so that both are convex and non-negative, with their global minimum at the reference state $\ualpha=\te 0$, $\beta=0$, respectively.

Since the thermodynamic forces are the negative gradients of the free energy, cf. Eq.~\eqref{eq:thermo_forces}, the gradient corrections give $B(0)=0$ and $\te A(\te{0})=\te{0}$. In the reference state of the hardening variables, the yield surface retains its initial size and position. The kinematic hardening force $\te A = -\partial \psi^\text{kin}/\partial \ualpha$ is obtained by differentiating Eq.~\eqref{eq:psi_kin_icnn} through the spectral decomposition \citep{lewis1996deri,miehe1993comp}. Convexity makes $\mathrm{d}\psi^\text{iso}/\mathrm{d}\beta$ non-decreasing from $\mathrm{d}\psi^\text{iso}/\mathrm{d}\beta\big|_{\beta=0}=0$, so that
\begin{equation}
B = -\frac{\mathrm{d}\psi^\text{iso}}{\mathrm{d}\beta} \leq 0 \; ,
\label{eq:B_sign}
\end{equation}
holds on the physical domain $\beta\geq0$ and isotropic hardening always expands the yield surface. Convexity of $\psi^\text{kin}$ plays the same role for kinematic hardening. For any fixed direction $\te d\in\Sym$, the resolved force $-\te A(\ualpha+t\,\te d):\te d$ is non-decreasing in $t$, so kinematic hardening maintains or increases the material's resistance to continued evolution in that direction. This is the architectural realization of the monotonic-hardening choice of Section~\ref{subsec:free_energy}, for both isotropic and kinematic hardening. Eq.~\eqref{eq:B_sign} specifically provides the sign $B\le0$ assumed in Section~\ref{subsec:evolution_thermo}.

\subsection{Flow potential}
\label{sec:flow_potential}

The flow potential of Eq.~\eqref{eq:G_full} is assembled from three learned components. Two of them, $\tilde\chi$ and $\chi^\text{kin}$, act on the eigenvalues of their respective tensor arguments, with isotropy enforced by permutation averaging (Section~\ref{sec:symmetry}). The third, $\chi^\text{iso}$, acts directly on the scalar $B$. Since the components enter additively, convexity of $\chi$ in $\tuple F=(\usig,\te A,B)$ follows directly from their individual properties. The effective stress $\usig^\text{eff}=\usig+\te A$ is linear in $(\usig,\te A)$, so $\tilde\chi(\usig^\text{eff})$ is convex in $(\usig,\te A)$ whenever it is convex in $\usig^\text{eff}$. $\chi^\text{kin}$ and $\chi^\text{iso}$ are each convex in their own arguments, and the remaining term $B$ is linear. It is this convexity in $\tuple F$ as a whole, rather than merely in each argument separately, that the thermodynamic argument of Appendix~\ref{app:thermo_consistency} requires.

The first component, i.e., $\tilde\chi$, is shared between the flow potential and the yield function. It is represented by a value-corrected ICNN, acting permutation-invariantly on the eigenvalues $(\sigma^\text{eff}_1,\sigma^\text{eff}_2,\sigma^\text{eff}_3)\in\SR^3$ of $\usig^\text{eff}$, i.e.,
\begin{equation}
\tilde{\chi}(\usig^\text{eff}) = \hat g_{\tilde\chi}\bigl(\sigma^\text{eff}_1(\usig^\text{eff}),\sigma^\text{eff}_2(\usig^\text{eff}),\sigma^\text{eff}_3(\usig^\text{eff})\bigr) - \hat g_{\tilde\chi}(0,0,0) \; , \qquad
\hat g_{\tilde\chi} := \mathcal{S\!P}\bigl(\bar{g}^{\mathrm{NN}}_{\tilde\chi}\bigr) \; .
\label{eq:G_icnn_corrected}
\end{equation}
Since the softplus is convex and non-decreasing, $\hat g_{\tilde\chi}$ remains convex and permutation-invariant, cf. Sections~\ref{sec:icnn} and \ref{sec:symmetry}. The outer softplus serves a separate purpose. It bounds $\hat g_{\tilde\chi}$ from below by zero, and hence $\tilde\chi$ from below by $-\hat g_{\tilde\chi}(0,0,0)$.
Unlike the other components, $\tilde\chi$ receives only a value correction, which provides the normalization $\tilde\chi(\te 0)=0$ of Section~\ref{subsec:evolution_thermo}, and no gradient correction. Its gradient at the origin is left free.

\begin{rmk}
    A gradient correction would remove the first-order dependence of $\tilde\chi$ on $\tr\usig^\text{eff}$ at the origin (Section~\ref{sec:zero_states}) and thereby its ability to represent a pressure-sensitive response such as Drucker--Prager plasticity. As a consequence, $\tilde\chi$ can be negative away from the yield surface. This is inconsequential, since thermodynamic consistency only requires $\chi\ge0$ on the yield surface itself, cf. Eq.~\eqref{eq:dissipation_condition_yield}. Non-negativity of $\chi$ on the yield surface is itself guaranteed by the parametrization of the yield function in Section~\ref{sec:yield_function} (Appendix~\ref{app:TC_non_associated}).
\end{rmk}

The second component, $\chi^\text{kin}$, is optional and extends $\partial\chi/\partial\te A$ beyond the contribution of $\tilde\chi$, thereby allowing the kinematic hardening variable to evolve non-associatively. It is constructed as a value- and gradient-corrected ICNN on the eigenvalues $(A_1,A_2,A_3)\in\SR^3$ of $\te A$, in the same manner as $\psi^\text{kin}$ in Eq.~\eqref{eq:psi_kin_icnn}, i.e.,
\begin{equation}
\chi^\text{kin}(\te A) = \bar{g}^{\mathrm{NN}}_{\chi^\text{kin}}\bigl(A_1(\te A),A_2(\te A),A_3(\te A)\bigr) - \bar{g}^{\mathrm{NN}}_{\chi^\text{kin}}(0,0,0) - \sum_{i=1}^{3}\left.\frac{\partial \bar{g}^{\mathrm{NN}}_{\chi^\text{kin}}}{\partial A_i}\right|_{\te A=\te 0}A_i(\te A) \; .
\label{eq:G_kin_icnn}
\end{equation}
In contrast to $\tilde\chi$, this component enters only the flow potential and not the yield function, so its sign is not controlled by the yield condition. Instead, convexity together with the value and gradient corrections places its global minimum at the origin with value zero, so that $\chi^\text{kin}\geq0$ and $\chi^\text{kin}(\te 0)=0$, as required in Section~\ref{subsec:non-associated}.

The third component, $\chi^\text{iso}$, is the scalar analog of $\chi^\text{kin}$ and is constructed in the same manner as $\psi^\text{iso}$ in Eq.~\eqref{eq:psi_iso_integral}, acting directly on $B$, i.e.,
\begin{equation}
\chi^\text{iso}(B) = g^{\mathrm{NN}}_{\chi^\text{iso}}(B) - g^{\mathrm{NN}}_{\chi^\text{iso}}(0) - \left.\frac{\mathrm{d}g^{\mathrm{NN}}_{\chi^\text{iso}}}{\mathrm{d}B}\right|_{B=0}B \; ,
\label{eq:chi_iso_integral}
\end{equation}
so that $\chi^\text{iso}\geq0$ and $\chi^\text{iso}(0)=0$ for the same reason. Like $\chi^\text{kin}$, it does not enter the yield function. It extends $\partial\chi/\partial B$ beyond the unit contribution of the linear term $B$ in Eq.~\eqref{eq:G_full}, so that the isotropic hardening rate $\dot\beta=\lambda\,(1+\mathrm{d}\chi^\text{iso}/\mathrm{d}B)$ may depart from the plastic multiplier $\lambda$.

\subsection{Yield function}
\label{sec:yield_function}

The yield function stated in Eq.~\eqref{eq:F_full_delta} realizes the additive parametrization of Section~\ref{subsec:non-associated}, Eq.~\eqref{eq:F_split}. The initial yield threshold $\sigma_{\text{y}_0}>0$ is a trainable parameter, kept positive through a softplus reparametrization, expressed as the softplus of an unconstrained raw parameter. In the associative case, no additional network is introduced. $\tilde\varphi\equiv\tilde\chi$, so that $\varphi(\usig^\text{eff}(\usig,\te A), B)=\tilde\chi(\usig^\text{eff}(\usig,\te A))+B-\sigma_{\text{y}_0}$, and the yield surface $\partial\set E$ is the level set $\tilde\chi(\usig^\text{eff}(\usig, \te A))=\sigma_{\text{y}_0}-B$ of $\tilde\chi$. In the non-associative case, $\Delta$ is parametrized as
\begin{equation}
\Delta(\usig^\text{eff}) = \sigma_{\text{y}_0}\Bigl[1-\mathcal{S\!P}\bigl(c - u(\usig^\text{eff})\bigr)\Bigr] \; , \qquad
u(\usig^\text{eff}) = \bar{g}^{\mathrm{NN}}_\Delta\bigl(\sigma^\text{eff}_1(\usig^\text{eff}),\sigma^\text{eff}_2(\usig^\text{eff}),\sigma^\text{eff}_3(\usig^\text{eff})\bigr) - \bar{g}^{\mathrm{NN}}_\Delta(0,0,0) \; ,
\label{eq:F_delta}
\end{equation}
where $\bar{g}^{\mathrm{NN}}_\Delta$ is an unconstrained multilayer perceptron, evaluated under the same permutation averaging as $\tilde\chi$ (Section~\ref{sec:symmetry}) so that $\tilde\varphi$ remains isotropic, and $c := \ln(e-1)$ is fixed such that $\mathcal{S\!P}(c)=1$. This is consistent with Section~\ref{subsec:non-associated}, which requires no convexity constraints on $\Delta$.

This construction has three properties. First, since the softplus is strictly positive, $\Delta(\usig^\text{eff})<\sigma_{\text{y}_0}$ holds for all $\usig^\text{eff}$ and every setting of the weights, which is the bound of Eq.~\eqref{eq:delta_bound}. Since $\Delta$ scales with $\sigma_{\text{y}_0}$, the bound follows the trainable yield threshold. It is unbounded from below, since $\mathcal{S\!P}(c-u)\to\infty$ as $u\to-\infty$. Second, the value correction of $u$, together with the choice of $c$, gives $\Delta(\te 0)=0$, so that $\varphi(\te 0,\te 0,0)=-\sigma_{\text{y}_0}$ as in the associative case, and $\sigma_{\text{y}_0}$ retains its meaning as the initial yield threshold. Third, $\Delta$ vanishes identically whenever $u\equiv0$, i.e., whenever $\bar g^\text{NN}_\Delta$ is constant, so that the associative model is nested within the non-associative parametrization (together with $\chi^\text{kin}\equiv\chi^\text{iso}\equiv0$, Section~\ref{subsec:associated}). Together with $\chi^\text{kin},\chi^\text{iso}\ge0$ (Section~\ref{sec:flow_potential}) and the convexity of $\chi$, the first property makes the architecture thermodynamically consistent by construction (Appendix~\ref{app:TC_non_associated}).

\subsection{Input scaling}
\label{sec:scaling}

All learned components above follow the same scaling convention, which is therefore stated once here. Each physical-unit argument, e.g., $\usig^\text{eff}$, $\te A$, $B$, $\ualpha$, or $\beta$, is first divided by a characteristic constant: a stress scale $\sigma_\text{s}>0$ for stress-like arguments and a strain scale $\varepsilon_\text{s}>0$ for strain-like arguments. This brings the network inputs into the unit range, which keeps the softplus activations in their non-saturated regime and the ICNNs well-conditioned for training. The origin-corrected network output is then multiplied by the scale matching its physical unit: $\sigma_\text{s}$ for the components of the flow potential, and $\sigma_\text{s}\varepsilon_\text{s}$ for the free energies $\psi^\text{iso}$ and $\psi^\text{kin}$. For $\Delta$, the output scale is already provided by the factor $\sigma_{\text{y}_0}$ in Eq.~\eqref{eq:F_delta}, so that only the input is scaled. Consequently, every function as written above takes and returns quantities in appropriate physical units. Since input and output are scaled by positive constants, convexity, isotropy, and the behavior at the zero states are unaffected. The specific scales used for each component are given in Appendix~\ref{app:scaling}.

\section{Training of the neural plasticity model}
\label{sec:algorithmic}

This section describes how the parameters of the neural plasticity model of Section~\ref{sec:neural_architecture} are identified from stress--strain data. These parameters are the weights and biases of the individual networks, denoted $\ve{\theta}_{\psi^\text{iso}}$, $\ve{\theta}_{\psi^\text{kin}}$, $\ve{\theta}_{\tilde\chi}$, $\ve{\theta}_{\chi^\text{iso}}$, $\ve{\theta}_{\chi^\text{kin}}$, and $\ve{\theta}_{\Delta}$, together with the initial yield threshold $\sigma_{\text{y}_0}$. We collect all of these in $\ve \theta$. Identifying $\ve\theta$ from stress--strain data is complicated by the fact that the internal variables are not observed, yet they are required to evaluate the stress.
Section~\ref{subsec:optimisation_problem} therefore formulates training as an optimization problem constrained by the return mapping and describes the computation of its gradients, and Section~\ref{sec:eig_degeneracy} addresses eigenvalue degeneracies arising from the spectral representation.

\subsection{Training formulation and optimization}
\label{subsec:optimisation_problem}

The dataset $\set D:=\left\{\set L_1,\ldots,\set L_{n^\text{load}}\right\}$ for calibration consists of $n^\text{load}\in\mathbb N$ load sequences of the form $\set L_l:=\left\{({}^{l,1} t, \ldots, {}^{l,n^\text{steps}} t ),\,({}^{l,1}\ueps, \ldots, {}^{l,n^\text{steps}}\ueps ),\,({}^{l,1}\usig, \ldots, {}^{l,n^\text{steps}}\usig )\right\}$, $l\in\{1,\ldots,n^\text{load}\}$, with $n^\text{steps}\in\mathbb N$ time steps each. As is common in machine learning, the dataset is split into a training set $\set D^\text{train}$, a validation set $\set D^\text{valid}$, and a test set $\set D^\text{test}$ \citep{herrmann2025deep}. Since the model is path-dependent, only entire load sequences are assigned to each split. The indices of the training sequences are collected in the set $\set T$. The validation set is used for early stopping only and does not enter the loss. Dataset sizes are reported for each example in Section~\ref{sec:numerical_results}.

Evaluating the stress for a prescribed strain sequence requires the internal variables ${}^{l,n}\tuple I$, which are unobserved in the data. We therefore determine the parameters by solving an optimization problem constrained by the time-discrete evolution equations. The constraints are the backward Euler discretization of the evolution law, Eq.~\eqref{eq:evolution:non_associated}, together with the KKT conditions, Eq.~\eqref{eq:KKT_conditions}, with $\gamma=\tau\lambda$. They thus constitute the time-discrete counterpart of a constraint given by ordinary differential equations with complementarity conditions.
The constrained optimization problem is defined as
\begin{align}
    \begin{split}
    &\hat{\ve \theta} = \underset{\ve \theta}{\arg \min}\;
    \frac{\dfrac{1}{6 |\set T| n^\text{steps}} \displaystyle\sum_{l\in \set{T}} \sum_{n=1}^{n^\text{steps}} \sum_{k=1}^{6}
    \Bigl(
    {}^{l,n}\sigma_k - \sigma_k\bigl({}^{l,n}\ueps,{}^{l,n}\tuple I(\ve \theta);\ve\theta\bigr)
    \Bigr)^2}{\sigma_\text{rms}^2} \; , \;
    \sigma_\text{rms}^2 := \frac{1}{6 |\set T| n^\text{steps}} \sum_{l\in \set{T}} \sum_{n=1}^{n^\text{steps}} \sum_{k=1}^{6} \bigl({}^{l,n}\sigma_k\bigr)^2
    \\
    &\text{s.t. } \;
    {}^{l,n}\tuple I - {}^{l,n-1}\tuple I = {}^{l,n}\gamma\,\diffp{\chi}{\tuple F}\bigg|_{\tuple F({}^{l,n}\ueps,{}^{l,n}\tuple I)} \; , \;\;
    {}^{l,n}\gamma \ge 0 \; \wedge \; {}^{l,n}\varphi\le 0 \; \wedge \; {}^{l,n}\gamma\,{}^{l,n}\varphi = 0 \; , \;\;
    {}^{l,0}\tuple I = (\te 0,\te 0,0) \; ,
    \end{split}
\label{eq:training_objective}
\end{align} 
for all $l\in\set T$ and $n\in\{1,\ldots,n^\text{steps}\}$. Here, $\sigma_k$, $k\in\{1,\ldots,6\}$, denote the six independent components of the stress tensor and  ${}^{l,n}\varphi:=\varphi\bigl(\tuple F({}^{l,n}\ueps,{}^{l,n}\tuple I)\bigr)$.
The loss $L(\ve\theta)$ is thus the mean squared error normalized by the magnitude of the training data.

At each time step, the constraints are solved by the predictor-corrector scheme of Section~\ref{subsec:return_mapping}. If the elastic predictor satisfies $\varphi\le0$, the step is elastic, with ${}^{l,n}\gamma=0$ and ${}^{l,n}\tuple I={}^{l,n-1}\tuple I$. Otherwise, the step is plastic, with ${}^{l,n}\gamma>0$ and ${}^{l,n}\varphi=0$, and ${}^{l,n}\tuple I$ and ${}^{l,n}\gamma$ solve
\begin{equation}
    \tuple R\bigl({}^{l,n}\tuple I,{}^{l,n}\gamma;\,{}^{l,n}\ueps,{}^{l,n-1}\tuple I,\ve\theta\bigr) = (\te 0, \te 0, 0, 0) \; ,
    \label{eq:training_constraint}
\end{equation}
i.e., Eqs.~\eqref{eq:return-map1} and~\eqref{eq:return-map2}. Since the constraints are solved for the current parameters $\ve\theta$ in every evaluation of the loss, they determine the internal variables as functions of $\ve\theta$, ${}^{l,n}\tuple I={}^{l,n}\tuple I(\ve\theta)$, without a closed-form expression, so the optimization reduces to one over $\ve\theta$ alone. This reduced formulation is common in the calibration of path-dependent classical \citep{seidl2021cali,kumar2025aco,akerson2025lear} as well as neural constitutive models \citep{fuhg2023modu,rosenkranz2024visc,meyer2023ther,kalina2026aph}.

The dependence of the internal variables on the parameters must, however, be accounted for in the gradient computation. 
To compute the sensitivities of the internal variables with respect to the parameters, $\mathrm{d}\,{}^{l,n}\tuple I/\mathrm{d}\ve\theta$, two strategies are available. The first treats the Newton iterations of Algorithm~\ref{ALG_RMA} as part of the computational graph and applies automatic differentiation to the entire unrolled iteration history. The second exploits that ${}^{l,n}\tuple I(\ve\theta)$ is defined implicitly by the converged residual, Eq.~\eqref{eq:training_constraint}. We use the second strategy, implemented in JAX. The return mapping is solved with the Newton solver of Optimistix \citep{rader2024opti}, which applies the implicit function theorem at the converged solution instead of differentiating through the Newton iterates themselves. The time stepping along each load path is implemented as a scan, and the gradient of the scalar loss with respect to $\ve\theta$ is obtained by reverse-mode automatic differentiation through it. This corresponds to an adjoint sensitivity analysis over time. 

\begin{figure}[tb]
    \centering
    \resizebox{\linewidth}{!}{\includegraphics{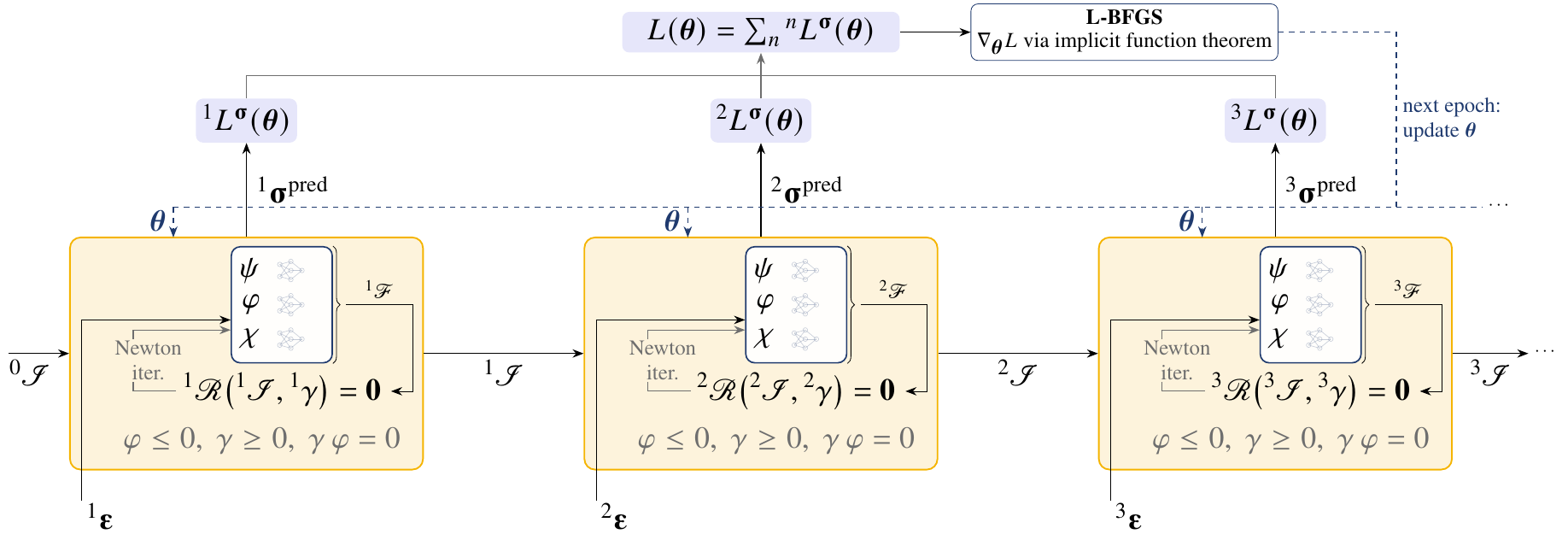}}
    \caption{Supervised training of the proposed neural plasticity model for a single load sequence. At each time step $n$, the learned free energy $\psi$ yields the thermodynamic forces ${}^{n}\tuple F:=({}^{n}\usig,{}^{n}\te A,{}^{n}B)$ (Eq.~\eqref{eq:thermo_forces}), on which the learned yield function $\varphi$ and flow potential $\chi$ act within the return-mapping residual ${}^{n}\tuple R$ (Eqs.~\eqref{eq:return-map1} and~\eqref{eq:return-map2}). The elastic predictor decides whether a step is plastic, consistent with the KKT conditions. In a plastic step, Newton's method solves ${}^{n}\tuple R=(\te 0,\te 0,0,0)$ for ${}^{n}\tuple I:=({}^{n}\ueps^\text{p},{}^{n}\ualpha,{}^{n}\beta)$ and ${}^{n}\gamma$. The predicted stress ${}^{n}\usig^\text{pred}(\ve\theta):=\usig\bigl({}^{n}\ueps,{}^{n}\tuple I(\ve\theta);\ve\theta\bigr)$ is the only supervised quantity. The gradient $\nabla_{\ve\theta}L(\ve \theta)$ of the loss is obtained by the implicit function theorem at each converged step and reverse-mode differentiation along the load sequence, and an L-BFGS step updates $\ve\theta$. The load index $l$ is omitted for brevity as only one sequence is considered.}
    \label{fig:training_loop}
\end{figure}

The loss is minimized with the L-BFGS method. The non-negativity of the ICNN weights required for convexity (Section~\ref{sec:icnn}) is already guaranteed by a softplus reparametrization, so no additional weight constraint needs to be enforced during optimization. A schematic representation of the training process for the case of a single time sequence is shown in Figure~\ref{fig:training_loop}.

\subsection{Differentiation at eigenvalue degeneracies}
\label{sec:eig_degeneracy}

Both the Newton iteration of Algorithm~\ref{ALG_RMA} and the gradient computation of Section~\ref{subsec:optimisation_problem} rely on the Jacobian $\tuple J=\partial\tuple R/\partial\tuple X$ of the return-mapping residual. Since $\tuple R$ contains the flow direction $\partial\chi/\partial\tuple F$ and the thermodynamic forces $\tuple F=-\partial\psi/\partial\tuple I$, $\tuple J$ contains second derivatives of the eigenvalue-based components $\tilde\chi$, $\psi^\text{kin}$, and $\chi^\text{kin}$ with respect to their tensor arguments. At degenerate states with repeated eigenvalues, a standard automatic-differentiation evaluation of these second derivatives yields non-finite values, although the exact derivatives are finite. This subsection explains the origin of this issue and the closed-form treatment used to avoid it. We describe the treatment for $\tilde\chi$, which applies equally to the other two components.

Let $\{\ve N_1,\ve N_2,\ve N_3\}$ be an orthonormal eigenbasis of $\usig^\text{eff}$, so that $\usig^\text{eff}=\sum_i\sigma^\text{eff}_i\,\ve N_i\otimes\ve N_i$. Such a basis always exists for a symmetric tensor, unique up to rotations within the corresponding eigenspace when eigenvalues repeat. The required Hessian reads
\begin{equation}
    \frac{\partial^2 \tilde\chi}{\partial \usig^\text{eff}\,\partial \usig^\text{eff}}
    =
    \sum_{i,j} \frac{\partial^2 \hat g_{\tilde\chi}}{\partial \sigma^\text{eff}_i\,\partial \sigma^\text{eff}_j}\,\te P_i\otimes\te P_j
    +
    \sum_{i \ne j} \vartheta_{ij}\,\te M_{ij}\otimes\te M_{ij} \; , \qquad
    \te P_i := \ve N_i\otimes\ve N_i \; , \quad
    \te M_{ij} := \tfrac{1}{2}(\ve N_i \otimes \ve N_j + \ve N_j \otimes \ve N_i) \; ,
    \label{eq:G_hessian_closed}
\end{equation}
with the divided-difference coefficient
\begin{equation}
\vartheta_{ij} = \frac{1}{\sigma^\text{eff}_i-\sigma^\text{eff}_j}\left(\diffp{\hat g_{\tilde\chi}}{\sigma^\text{eff}_i}-\diffp{\hat g_{\tilde\chi}}{\sigma^\text{eff}_j}\right)
\label{eq:theta_divided}
\end{equation}
\citep{miehe1993comp,souzaneto2008comp}; see Appendix~\ref{app:eig_derivatives} for the derivation. Since $\hat g_{\tilde\chi}$ is smooth and permutation-invariant, $\tilde\chi$ is twice continuously differentiable in $\usig^\text{eff}$ everywhere \citep{lewis2001twic}, including at degenerate states with repeated eigenvalues of $\usig^\text{eff}$, i.e., states with two equal eigenvalues, such as uniaxial and equibiaxial stress, and hydrostatic states. Correspondingly, $\vartheta_{ij}$ remains finite as $\sigma^\text{eff}_j\to\sigma^\text{eff}_i$, since by permutation invariance the numerator and denominator vanish together, with the limit given by $\partial^2\hat g_{\tilde\chi}/\partial(\sigma^\text{eff}_i)^2-\partial^2\hat g_{\tilde\chi}/\partial\sigma^\text{eff}_i\partial\sigma^\text{eff}_j$. This limit takes the same value for all pairs $i\ne j$ within a degenerate eigenspace, so that Eq.~\eqref{eq:G_hessian_closed} does not depend on the choice of the eigenbasis.

A standard automatic-differentiation evaluation does not exploit this cancellation. It differentiates through the eigendecomposition, whose eigenvector derivatives contain factors $1/(\sigma^\text{eff}_i-\sigma^\text{eff}_j)$ \citep{liao2019diff}, and multiplies them by the derivatives of $\hat g_{\tilde\chi}$ only afterwards. Although the exact Hessian is finite, the resulting Jacobian therefore contains non-finite entries at every degenerate state, and the Newton iteration of Algorithm~\ref{ALG_RMA} breaks down there. We avoid this by performing the spectral decomposition outside the network and registering a custom derivative rule that returns Eq.~\eqref{eq:G_hessian_closed} directly. The derivatives of $\hat g_{\tilde\chi}$ with respect to the eigenvalues are obtained by standard automatic differentiation of a scalar function of three scalars, which is unproblematic. The coefficient $\vartheta_{ij}$ is evaluated as
\begin{equation}
\vartheta_{ij} =
\begin{cases}
\dfrac{1}{\sigma^\text{eff}_i-\sigma^\text{eff}_j}\left(\diffp{\hat g_{\tilde\chi}}{\sigma^\text{eff}_i}-\diffp{\hat g_{\tilde\chi}}{\sigma^\text{eff}_j}\right) \; , & |\sigma^\text{eff}_i-\sigma^\text{eff}_j| > \tau_\text{deg} \; ,\\[3mm]
\dfrac{\partial^2\hat g_{\tilde\chi}}{\partial(\sigma^\text{eff}_i)^2}-\dfrac{\partial^2\hat g_{\tilde\chi}}{\partial\sigma^\text{eff}_i\,\partial\sigma^\text{eff}_j} \; , & |\sigma^\text{eff}_i-\sigma^\text{eff}_j| \le \tau_\text{deg} \; ,
\end{cases}
\label{eq:theta_branch}
\end{equation}
with the threshold $\tau_\text{deg}=10^{-10}$ applied to the scaled eigenvalues (Section~\ref{sec:scaling}) \citep{miehe1993comp,souzaneto2008comp}. The second branch is exact at coincidence, and both branches introduce only small errors in the vicinity of the threshold, which are negligible for the Newton iteration. The Newton iteration therefore retains its quadratic convergence in practice, including at degenerate stress states. For comparison, Appendix~\ref{app:eig_material_point_I1_I2} also considers a Lorentzian regularization of $\vartheta_{ij}$ \citep{liao2019diff}, derived in Appendix~\ref{app:eig_derivatives}.

\section{Results}
\label{sec:numerical_results}

In this section, we present calibration studies that examine model performance across interpolation, fit quality within training coverage, and extrapolation beyond it. The data generation procedures are described in detail in Appendix~\ref{sec:data_generation} (Table~\ref{tab:data_sources}). Specifically, we structure our evaluation around three key objectives:

\begin{enumerate}[label=(\roman*)]
    \item We demonstrate that the model recovers known constitutive behavior when ground truth is available by isolating the yield criterion and flow rule, the isotropic-hardening law, and the kinematic-hardening law individually (Section~\ref{sec:single_mechanism_studies}).
    \item We quantify the improvements gained by enforcing isotropy over an otherwise identical triclinic baseline, using specimen data for a Lode-sensitive material (Section~\ref{sec:plane_stress_symmetry_ablation}).
    \item We evaluate predictive performance on RVE-simulated data whose ground truth is not known in closed form, using an open-cell foam RVE case study (Section~\ref{sec:axis4_flow_rule}).
\end{enumerate}

We report dataset sizes using the number of training load paths $n^\text{load}$ and timesteps per path $n^\text{steps}$, noting any exceptions where applicable. Appendix~\ref{sec:data_sources} provides complete descriptions of all data sources.
We report two additional comparisons on material-patch data: a second isotropy-enforcement study and a comparison against a generic sequence-model baseline (Appendix~\ref{sec:material_patch_ablations}).

All models are trained on full stress-strain load-path sequences by minimizing the objective~\eqref{eq:training_objective} as described in Section~\ref{subsec:optimisation_problem}. The implementation has been done in JAX \citep{bradbury2018jax} with Equinox \citep{kidger2021equi} for the model and Optax \citep{deepmind2020the} for the optimizer (L-BFGS throughout); the return-mapping Newton solve is provided by Optimistix \citep{rader2024opti}, differentiated via the implicit-function-theorem approach described in Section~\ref{subsec:optimisation_problem}. Training runs on CPU only (48-core Intel Xeon Cascade Lake nodes), since this solve is a sequential, float64-precision computation that a GPU does not accelerate; it took from approximately 5.6 s/epoch for the smallest study to approximately 21 s/epoch for the largest (the open-cell foam RVE dataset). Training ran for 500 to roughly 3000 epochs depending on the study, with $\set D^\text{valid}$ monitored throughout for early stopping. Every learned component in Table~\ref{tab:design_summary} is parametrized identically across all studies below, each a small network with a width of 16 units and 2 hidden layers; the resulting total parameter count per study, summing every active component, ranges from roughly 350 (a single yield/flow-potential network) up to roughly 1300 (the open-cell foam RVE study, which combines a non-associative yield correction with isotropic hardening).

\begin{rmk}[Reporting conventions]
\label{rmk:reporting_conventions}
Unless a table states otherwise, we report fit quality using two metrics: the coefficient of determination $R^2$ and a normalized root-mean-square error (NRMSE). We compute both metrics per test path and per stress component as the time-series RMSE divided by its own mean absolute stress value, averaged over the three normal Mandel components across all test paths (excluding shear components by default). The coefficient $R^2$ is reported to 4 significant figures. When compared values are close enough to 1 that standard rounding conceals differences, we report the complement $1-R^2$ in scientific notation. We report NRMSE as a percentage to 3 significant figures, or in scientific notation for values below $0.001\%$. Following \citet{ferreira2025auto}, we define NRMSE as the ratio of a stress component's root-mean-square error (RMSE) to its mean absolute value (MAV), both evaluated over the path's time series using ground-truth stress:
\begin{equation}
\mathrm{NRMSE} = \mathrm{RMSE}/\mathrm{MAV} \,,
\label{eq:NRMSE_def}
\end{equation}
averaged according to the convention above.
\end{rmk}

\begin{table}[b]
    \centering
    \begin{minipage}{0.48\textwidth}
        \centering
        \small
        \caption{Closed-form (ideal) plasticity models.}
        \label{tab:analytical_yield_functions}
        \begin{tabular}{ll}
            \toprule
            Model & Yield function / flow potential \\
            \midrule
            Von Mises & $\tilde\varphi = \sqrt{3J_2} - (\sigma_{y_0} - B)$ \\
            Drucker--Prager & $\tilde\varphi = \sqrt{J_2} + M_\varphi(I_1 - \kappa)$ \\
             & $\tilde\chi = \sqrt{J_2} + M_\chi(I_1 - \kappa), \ M_\chi\le M_\varphi$ \\
            LZY & $\tilde\varphi = a\big(bI_1 + \sqrt[3]{\sqrt{J_2^3 - cJ_3^2} - dJ_3}\big)$ \\
             & \quad ~~$- (\sigma_{y_0}-B)$ \\
            \bottomrule
        \end{tabular}
    \end{minipage}%
    \hfill
    \begin{minipage}{0.48\textwidth}
        \centering
        \small
        \caption{Performance metrics: ideal plasticity datasets.}
        \label{tab:axis1_summary}
        \begin{tabular}{lcc}
            \toprule
            Ground truth & Test $1-R^2$ & Test NRMSE \\
            \midrule
            Von Mises & $2.60\times10^{-13}$ & $1.00\times10^{-4}\%$ \\[1ex]
            Drucker--Prager (assoc.) & $7.00\times10^{-4}$ & 1.87\% \\[1ex]
            Drucker--Prager (non-assoc.) & $5.00\times10^{-4}$ & 1.86\% \\[1ex]
            LZY & $4.48\times10^{-5}$ & 0.679\% \\[2.4ex]
            \bottomrule
        \end{tabular}
    \end{minipage}
\end{table}

\subsection{Demonstration of model expressivity via known constitutive laws}
\label{sec:single_mechanism_studies}

In the following three studies, we examine the framework's expressivity by isolating individual mechanisms (the yield criterion and flow rule, the isotropic-hardening law, and the kinematic-hardening law) in turn, varying one mechanism at a time while holding the others fixed. We train the learned architecture end-to-end from full stress-strain load-path data through the return map as described in Section~\ref{subsec:optimisation_problem}, without applying direct supervision to internal potentials or hardening variables. The flow potential is represented by an ICNN $\tilde\chi$, with $\tilde\varphi \equiv \tilde\chi$ in associative settings. Additional ICNNs provide the isotropic and kinematic hardening modeling components as applicable. All three studies use a synthetic dataset of 32 training paths with 200 timesteps per path, generated from known constitutive laws (Appendix~\ref{sec:data_sources}, source 1).

\subsubsection{Ideal plasticity}
\label{sec:axis1_yield_flexibility}

\begin{figure}[t]
    \centering
    \includegraphics{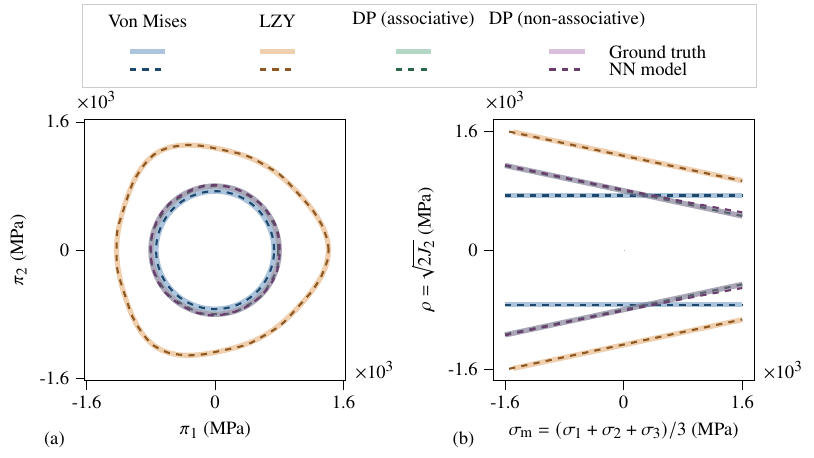}
    \caption{(a) $\pi$-plane cross-section ($\sigma_\text{m}=0$) and (b) meridian cross-section (shear meridian, $\theta_L=0$) of the learned yield surface for four ideal (no-hardening) reference materials. For each material, the thick translucent band is its ground truth and the darker dashed line is the model prediction, both in the material's own color.}
    \label{fig:axis1_pi_plane}
\end{figure}

Here we want to confirm that our neural plasticity model captures qualitatively different yield-surface geometries through parameter fitting without changing the network structure.
To do so, we train the network representation independently on four ideal (non-hardening) ground-truth plasticity laws: Von Mises \citep{mises1913mech}, associative and non-associative Drucker--Prager \citep{drucker1952soil,simo1998comp,souzaneto2008comp,lambrecht2001ano}, and LZY \citep{lou2022age}. The expressions for the ground-truth yield criteria are provided in Table~\ref{tab:analytical_yield_functions} in terms of the principal invariants $I_1,I_2,I_3$ of $\usig^\text{eff}$ and the deviatoric invariants $J_2 = \tfrac13 I_1^2 - I_2$ and $J_3 = \tfrac{2}{27}I_1^3 - \tfrac13 I_1 I_2 + I_3$, setting $\kappa \coloneqq (\sigma_{y_0}-B)/M_\varphi$ for Drucker--Prager plasticity. Associative behavior follows from setting $\tilde\chi\equiv\tilde\varphi$ directly for Von Mises and LZY, or $M_\chi=M_\varphi$ for Drucker--Prager. Parameter values for all three laws are given in Appendix~\ref{sec:source_analytical}.

For fitting the proposed neural plasticity model to non-associative Drucker--Prager plasticity, we add the yield-function correction described in Section~\ref{sec:yield_function}, i.e., $\tilde\varphi = \tilde\chi + \Delta$ (Eq.~\eqref{eq:F_delta}), where a secondary unconstrained network outputs $\Delta$ without altering the flow potential $\tilde\chi$.

As summarized in Table~\ref{tab:axis1_summary}, the trainings of the neural model results in a close-to-exact fit for all four cases on their own held-out test data. Figure~\ref{fig:axis1_pi_plane} compares the learned yield surface against ground truths via two cross-sections. Panel (a) cuts the yield surface at fixed hydrostatic stress $\sigma_\text{m}=(\sigma_1+\sigma_2+\sigma_3)/3$ (the $\pi$-plane), with axes $\pi_1,\pi_2$ spanning that plane. Panel (b) cuts the yield surface at the shear meridian ($\theta_L=0$), with axes $\sigma_\text{m}$ and $\rho=\sqrt{2J_2}$. The trained models recover circular, Lode-angle-independent cross-sections for Von Mises and Drucker--Prager, and capture the non-circular, Lode-angle-dependent cross-sections arising from LZY's cube-root yield-surface formulation.

\subsubsection{Plasticity with isotropic hardening}
\label{sec:axis2_iso_hardening}

Following our assessment of ideal plasticity, we evaluate how well the framework can learn different isotropic hardening behaviors directly from stress-strain data. To that end, we generate synthetic datasets using the Von Mises yield criterion while varying the isotropic-hardening law, as given in Table~\ref{tab:analytical_iso_hardening_laws}: linear, power-law \citep{ludwik1909elem}, and saturating \citep{voce1948the}. Appendix~\ref{sec:source_analytical} provides the associated parameters.
\begin{table}[tb]
    \centering
    \begin{minipage}[t]{0.45\textwidth}
        \centering
        \small
        \caption{Ground-truth isotropic hardening laws.}
        \label{tab:analytical_iso_hardening_laws}
        \begin{tabular}{ll}
            \toprule
            Law & Current yield stress $\sigma_y(\beta)$ \\
            \midrule
            Linear & $\sigma_{y_0} + H\beta$ \\
            Power-law (Ludwik) & $\sigma_{y_0} + s_1\beta^{s_2}$ \\
            Voce (saturating) & $\sigma_{y_0} + Q\big(1-e^{-b\beta}\big)$ \\
            \bottomrule
        \end{tabular}
    \end{minipage}%
    \hfill
    \begin{minipage}[t]{0.55\textwidth}
        \centering
        \small
        \caption{Performance metrics: Von Mises isotropic hardening datasets.}
        \label{tab:axis2_summary}
        \begin{tabular}{lcc}
            \toprule
            Ground truth law & Test $1-R^2$ & Test NRMSE \\
            \midrule
            Linear & $2.63\times10^{-7}$ & 0.0299\% \\
            Power-law (Ludwik) & $6.94\times10^{-7}$ & 0.0888\% \\
            Voce (saturating) & $1.74\times10^{-7}$ & 0.0328\% \\
            \bottomrule
        \end{tabular}
    \end{minipage}
\end{table}

\begin{figure}[tb] 
    \centering
    \includegraphics{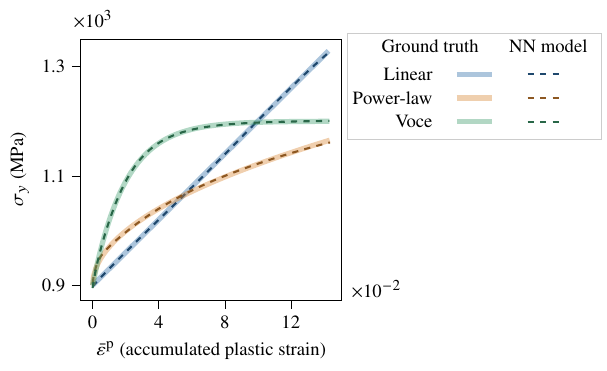}
    \caption{Current yield stress $\sigma_\text{y}$ vs.\ accumulated plastic strain $\bar\varepsilon^{\mathrm{p}}$, three isotropic hardening laws (Von Mises). For each law, the thick translucent band is its ground truth and the darker dashed line is the model prediction, both in the law's own color.}
    \label{fig:axis2_hardening_curve}
\end{figure}

Since the training constrains only the mapping from strain to stress, each network parametrizes its internal hardening variable $\beta$ on an arbitrary numerical scale. Comparing values of $\beta$ across learned models or against each ground-truth law's own definition is therefore not physically meaningful. Instead, we evaluate performance using observable physical quantities derived from predicted trajectories, specifically the current yield stress $\sigma_\text{y}$ and accumulated plastic strain $\bar\varepsilon^{\mathrm{p}}$.
Figure~\ref{fig:axis2_hardening_curve} compares the courses $\sigma_\text{y}$ over $\bar\varepsilon^{\mathrm{p}}$ against their ground-truth references. The learned models capture the distinct qualitative trends of each law: unbounded linear growth, asymptotic Voce saturation, and power-law growth with a continuously decreasing hardening rate.
As summarized in Table~\ref{tab:axis2_summary}, we observe close convergence of the learned models to the respective ground-truth laws across the three cases.
\subsubsection{Plasticity with kinematic hardening}
\label{sec:axis3_kin_hardening}

We next evaluate kinematic-hardening expressivity of our model by benchmarking its capability in capturing linear \citep{prager1949rece} and Armstrong--Frederick \citep{armstrong1966ama} kinematic-hardening behaviors directly from stress-strain data. Similar to the previous study, we maintain the Von Mises yield criterion and employ the kinematic hardening laws given in Table~\ref{tab:analytical_kin_hardening_laws} to generate the datasets. The neural plasticity models trained on each of these ground-truth laws achieve high test set accuracy (Table~\ref{tab:axis3_summary}). Appendix~\ref{sec:source_analytical} lists the corresponding parameters of the ground-truth laws.

\begin{table}[tb]
    \centering
    \begin{minipage}[t]{0.45\textwidth}
        \centering
        \small
        \caption{Analytical kinematic hardening laws.}
        \label{tab:analytical_kin_hardening_laws}
        \begin{tabular}{lll}
            \toprule
            Law & $\psi^\mathrm{kin}(\ualpha)$ & $\chi^\mathrm{kin}(\te A)$ \\
            \midrule
            \vspace{0.2ex}
            Linear (Prager) & $\tfrac12 H\,\ualpha\!:\!\ualpha$ & $0$ \\[1ex]
            Armstrong--Frederick & $\tfrac12 H\,\ualpha\!:\!\ualpha$ & $\tfrac{\gamma_\mathrm{af}}{2H}\,\te A\!:\!\te A$ \\
            \bottomrule
        \end{tabular}
    \end{minipage}%
    \hfill
    \begin{minipage}[t]{0.55\textwidth}
        \centering
        \small
        \caption{Performance metrics: Von Mises kinematic hardening datasets.}
        \label{tab:axis3_summary}
        \begin{tabular}{lcc}
            \toprule
            Ground truth law & Test $1-R^2$ & Test NRMSE \\
            \midrule
            Linear (Prager) & $4.29\times10^{-7}$ & 0.0769\% \\[1ex]
            Armstrong--Frederick & $1.37\times10^{-7}$ & 0.0940\% \\[1ex]
            \bottomrule
        \end{tabular}
    \end{minipage}
\end{table}

\begin{figure}[tb]
    \centering
    \includegraphics{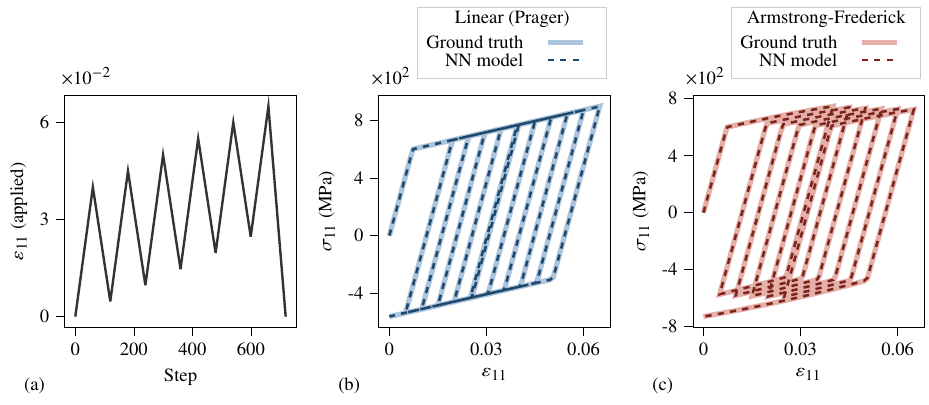}
    \caption{(a) Applied cyclic loading path used to compare the two kinematic hardening laws beyond their own training data, an escalating asymmetric strain cycle, each successive cycle's peak strain increasing by a fixed increment. (b) Stress response for linear (Prager) kinematic hardening. (c) Stress response for Armstrong--Frederick. In (b),(c), the thick translucent band is the ground truth's response and the darker dashed line is the neural model prediction.}
    \label{fig:axis3_strain_input}
    \label{fig:axis3_linear}
\end{figure}

We further evaluate both trained neural models on an escalating, asymmetric cyclic strain path (not part of the training set) to give a clear visual picture of the two kinematic-hardening laws and how well the trained models capture each (Figure~\ref{fig:axis3_strain_input}a). Figure~\ref{fig:axis3_linear}(b),(c) plots the predicted stress responses of the trained models alongside the ground-truth solutions. It highlights the expressivity of the proposed framework in capturing the underlying physics of the reference kinematic hardening laws, specifically the continuing back-stress accumulation of the linear kinematic hardening law, as well as the back-stress saturation of the Armstrong--Frederick law.

\subsection{Isotropy enforcement under plane stress}
\label{sec:plane_stress_symmetry_ablation}

\begin{figure}[t]
    \centering
    \begin{tabular}{c@{\hspace{3em}}c@{\hspace{3em}}c}
    {\includegraphics[height=3.4cm]{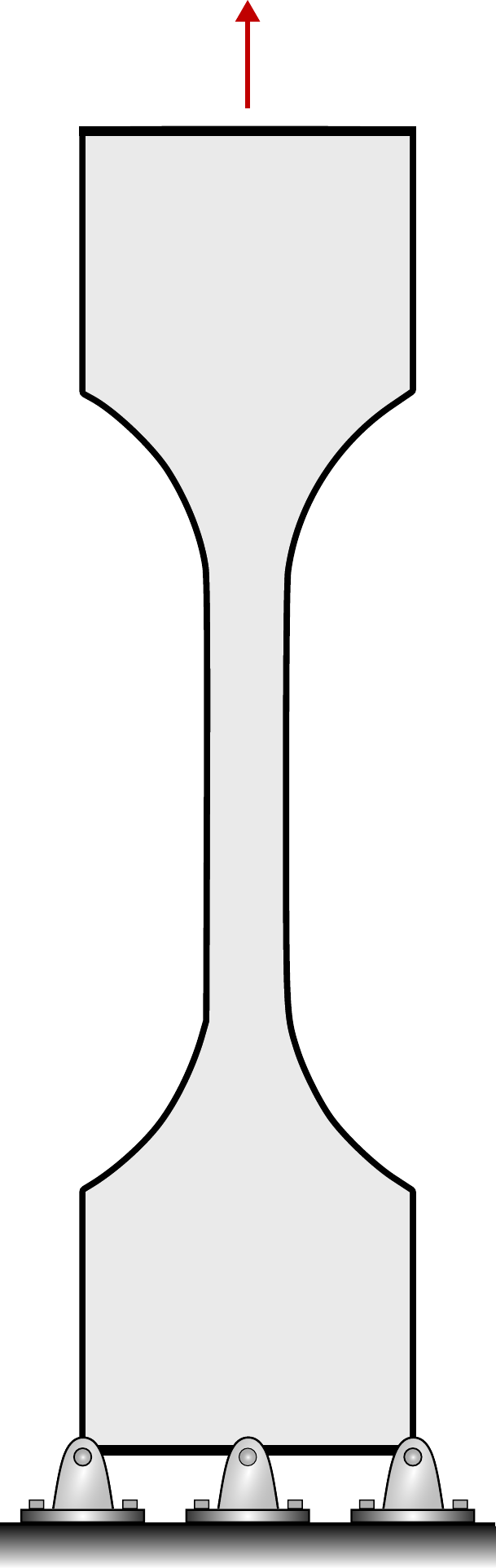}} &
    {\includegraphics[height=3.4cm]{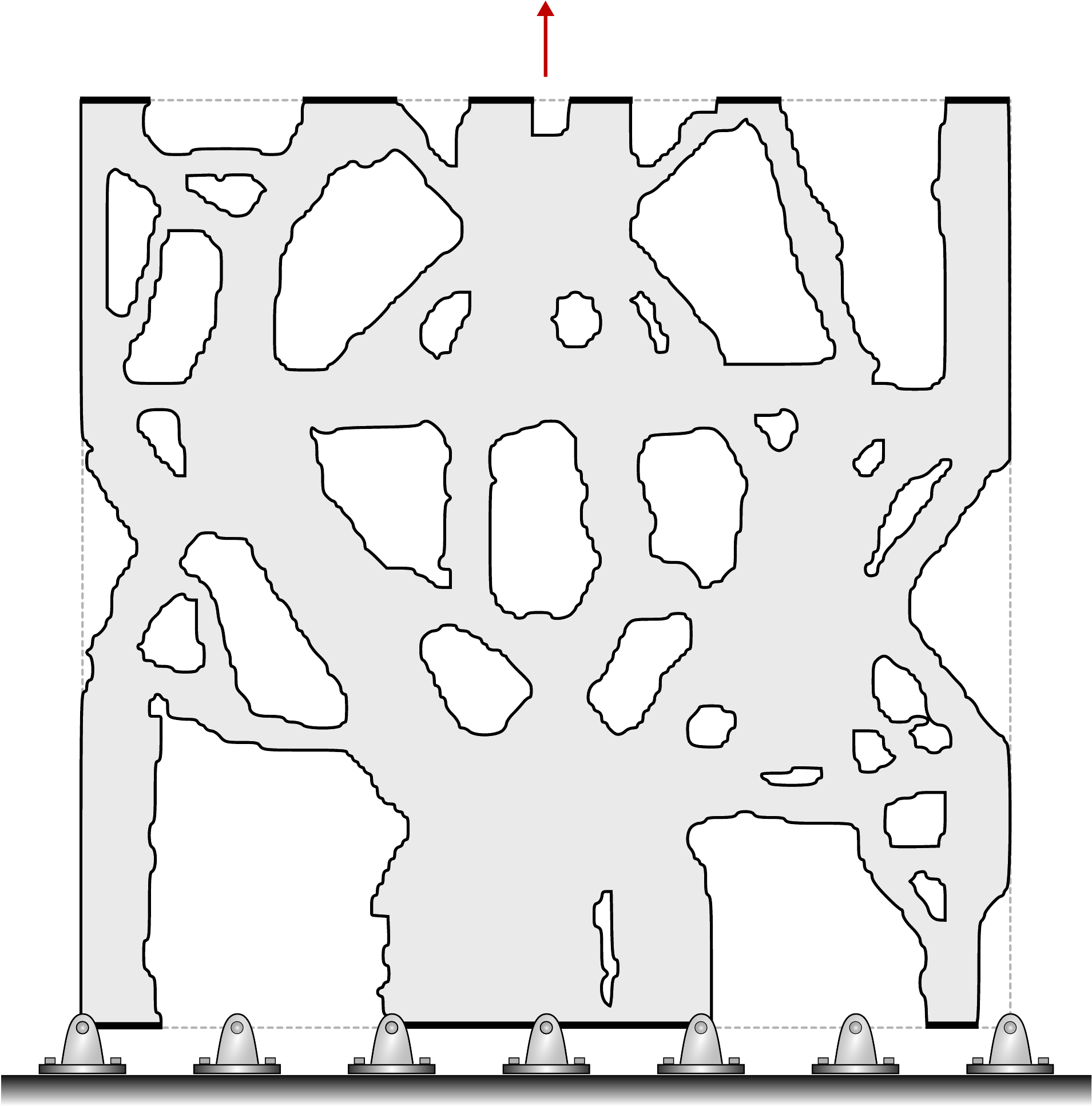}} &
    {\includegraphics[height=3.4cm]{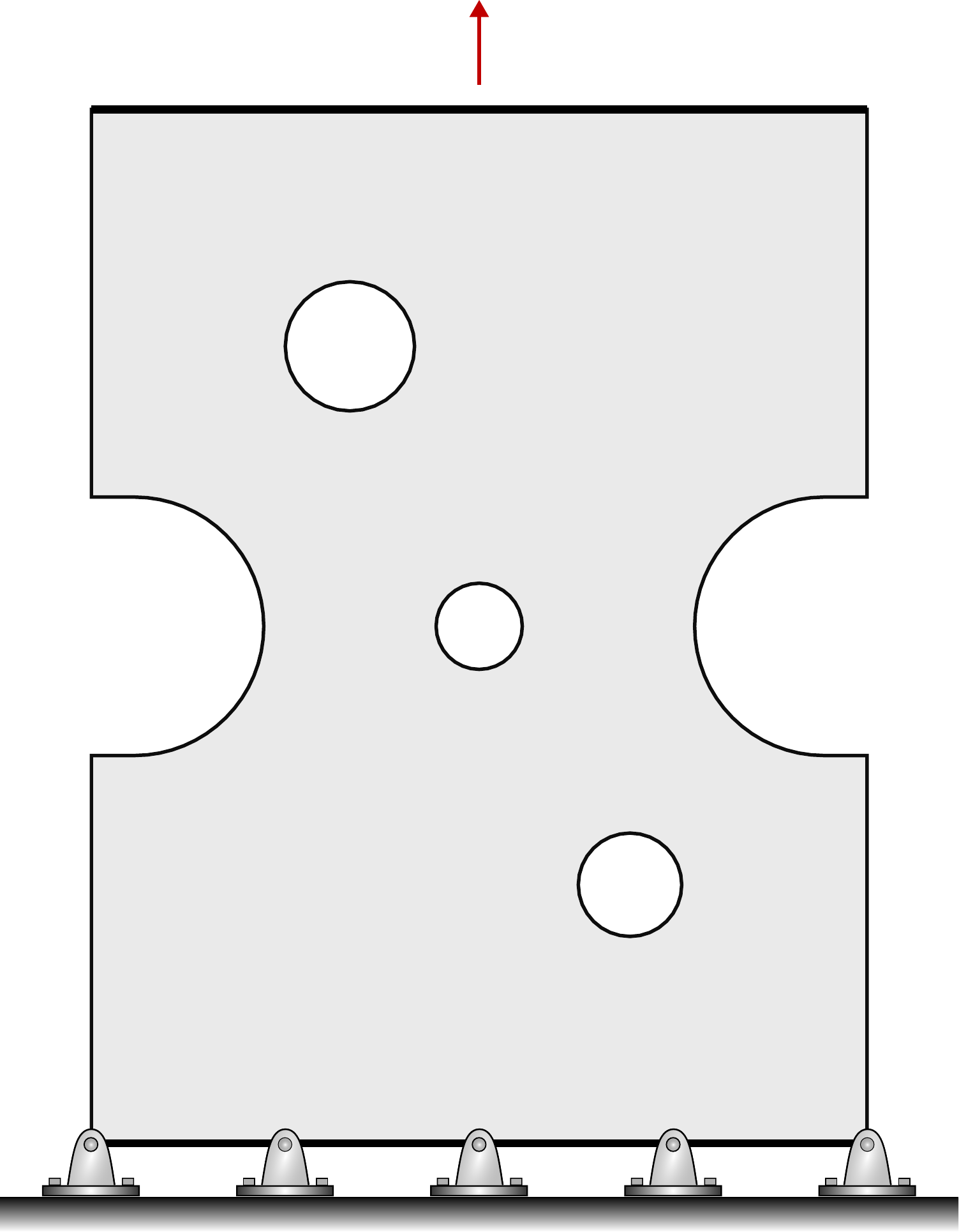}} \\
    {\small (a) Dogbone} &{\small (b) Topology-optimized } & {\small (c) FEM validation geometry}\\
    \end{tabular}
    \caption{(a) Standard dogbone specimen, and (b) topology-optimized specimen designed to elicit rich strain-path diversity under uniaxial loading. (c) Unseen geometry used for independent FEM validation: a perforated 18$\times$24~mm plate with two opposing notches and three circular holes of different radii, loaded uniaxially at the top edge with the bottom edge fixed.}
    \label{fig:dogbone_to_geometry}
\end{figure}

\begin{figure}[t]
    \centering
    \includegraphics{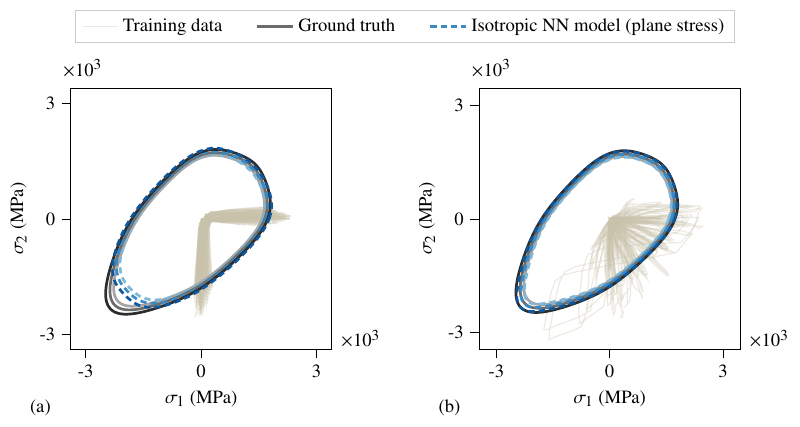}

    \vspace{0.2cm}
    \includegraphics{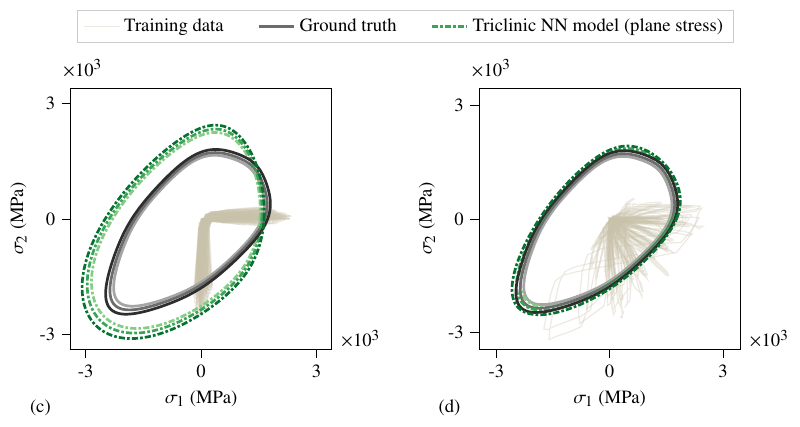}
    \caption{Comparison of the true yield locus against the isotropic NN model (a-b) and the triclinic NN model (c-d) in the $(\sigma_1,\sigma_2)$ plane, for dogbone (a, c) and topology-optimized (b, d) specimens, at three matched hardening levels each. Hardening levels are matched across models using $\rho=\sqrt{2J_2}$ evaluated along the uniaxial-tension direction ($\rho=1420$, $1480$, $1550$~MPa; shade = level, lightest to darkest; legend shows one representative shade per curve family). Each specimen's training loading paths, diagonalized to the same principal-stress axes, are traced underneath (tan).}
    \label{fig:plane_stress_sigma1_sigma2}
\end{figure}

Having established that the modeling framework can recover known constitutive behavior, we now assess how well our models learn when presented with less diverse datasets. Such scenarios arise in practice when the material is tested using standard specimen geometries, which may not produce a sufficiently rich set of stress-strain paths to fully inform the modeling framework. To this end, we employ synthetic stress-strain datasets extracted from two different specimen geometries shown in Figure~\ref{fig:dogbone_to_geometry}(a,b): a standard dogbone specimen and a topology-optimized specimen designed to elicit rich strain-path diversity under uniaxial loading. The topology-optimized geometry is designed using the methodology of \citet{yin2026desi}, which optimizes the internal ligaments of the geometry to transform an applied uniaxial displacement into multi-axial, non-proportional strain paths. This generates high strain-path diversity within a single test setup, in contrast to the dogbone specimen, which produces a narrow band of stress-strain paths.\footnote{The topology-optimized specimen is particularly interesting for unsupervised training of neural constitutive models based on full-field data, e.g., via finite element model updating or the equilibrium gap method. Here, however, we merely demonstrate how the neural plasticity model generally performs with respect to the information contained in the specimen and calibrate it directly using stress-strain data.}
The ground-truth material law governing both specimens corresponds to an LZY yield surface with power-law isotropic hardening (Tables~\ref{tab:analytical_yield_functions} and~\ref{tab:analytical_iso_hardening_laws}). Both specimens are simulated using finite element analysis and subjected to a cyclic loading-unloading displacement. The resulting local (material-point-level) stress-strain paths are then selected to build the training dataset. To keep the dataset tractable, it is filtered to contain loading paths with non-trivial plastic content. Note that, since the dataset is generated via finite element simulation, the ground-truth reference to which the model is compared is still the generating law.

We also use this study to evaluate the effect of enforcing isotropy on the model's generalization abilities. The specimens considered are subject to plane stress conditions, and as a result the out-of-plane stresses are set to zero. From a material-model perspective, isotropy then corresponds to a reflection symmetry. We compare this isotropy-enforcing representation against a triclinic baseline. The triclinic neural plasticity model instead uses the component-based parametrization of Remark~\ref{remark:triclinic} (Eq.~\eqref{eq:triclinic_component}) for every sub-potential, in place of the eigenvalue-based parametrization of Section~\ref{sec:symmetry}.

\begin{table}[tb]
    \centering
    \small
    \caption{Performance metrics: isotropy enforcement under plane stress.}
    \label{tab:plane_stress_summary}
    \begin{tabular}{lcccc}
        \toprule
        & \multicolumn{2}{c}{Isotropic NN model} & \multicolumn{2}{c}{Triclinic NN model} \\
        \cmidrule(lr){2-3} \cmidrule(lr){4-5}
        Specimen & $R^2$ & NRMSE & $R^2$ & NRMSE \\
        \midrule
        Dogbone & 0.9815 & 11.03\% & 0.9650 & 88.85\% \\
        Topology-optimized & 0.9999 & 1.30\% & 0.9990 & 20.32\% \\
        \bottomrule
    \end{tabular}
\end{table}

\begin{figure}[t]
    \centering
    \includegraphics{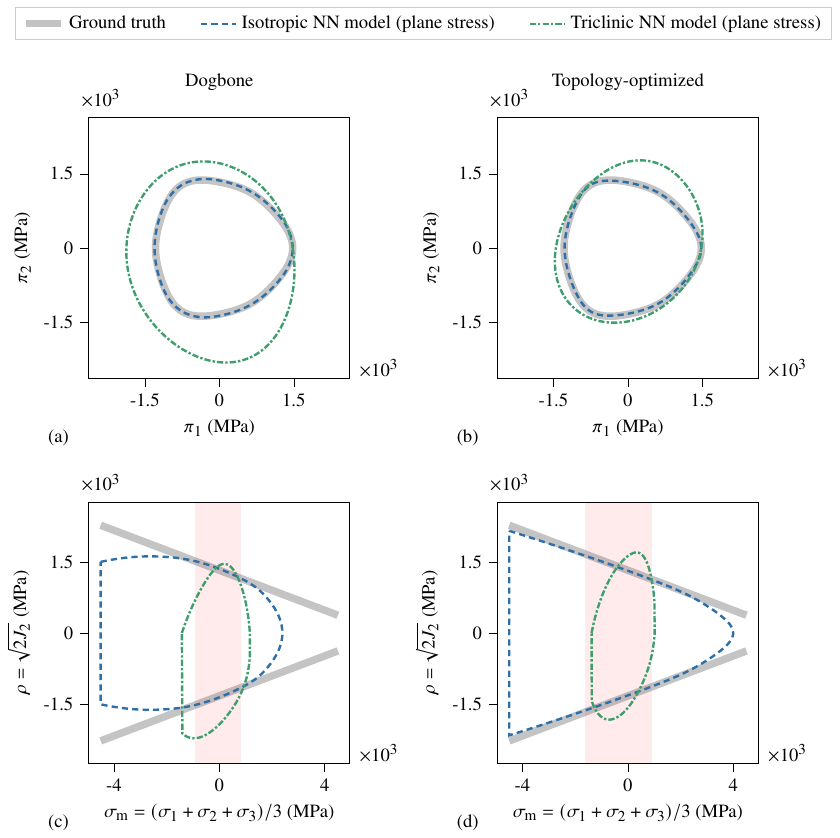}
    \caption{Dogbone (left column) vs.\ topology-optimized (right column) specimen, true material vs.\ isotropic NN model vs.\ triclinic NN model. (a)-(b) $\pi$-plane cross-section at $\sigma_\text{m}=0$, within training coverage, evaluating generalization performance. (c)-(d) Meridian cross-section, full $\pm4500$~MPa sweep extending well beyond training coverage, evaluating extrapolation behavior; shaded band marks each specimen's own training $\sigma_\text{m}$ coverage.}
    \label{fig:plane_stress_merged}
\end{figure}

Figure~\ref{fig:plane_stress_sigma1_sigma2} demonstrates how well the isotropic and triclinic models capture the yield surface in the $(\sigma_1,\sigma_2)$ plane. Because the two representations parameterize hardening differently, the three plotted levels are matched across models by $\rho=\sqrt{2J_2}$ along the uniaxial-tension direction. The overlaid training paths, diagonalized to the same principal-stress axes, indicate each specimen's stress-space coverage. The training paths generated from dogbone specimen remain concentrated in two narrow bands, while training paths from the topology-optimized specimen cover a broader domain extending closer to the outer boundary of the yield locus. Within training coverage, the isotropic models trained on both specimens accurately match the reference loci across all hardening levels. The triclinic model trained on the topology-optimized specimen closely matches the reference loci as well, but the triclinic model trained on the dogbone specimen visibly deviates.

\paragraph{Generalization performance.}
We evaluate generalization performance using an independent test set consisting of random stress-strain paths generated directly from the ground-truth LZY law. The isotropic models achieve high accuracy, with NRMSE values of 11.03\% for the model trained on the dogbone specimen and 1.30\% for the model trained on the topology-optimized specimen (Table~\ref{tab:plane_stress_summary}). The triclinic models reach higher errors of 88.85\% and 20.32\%, respectively, for the same two specimens. Enforcing isotropy reduces prediction error by factors of 8.06 and 15.58 on the dogbone and topology-optimized datasets. The isotropic models also achieve higher $R^2$ values than their triclinic counterparts for both specimens (Table~\ref{tab:plane_stress_summary}). These two metrics normalize errors differently, one by the overall test-set variance, the other by each path's own magnitude, so they need not agree when compared across specimens. Within each specimen, however, both consistently favor the isotropic model. NRMSE values in Table~\ref{tab:plane_stress_summary} average over the three in-plane stress components ($\sigma_{11}, \sigma_{22}, \sigma_{12}$), since a meaningful comparison for the non-isotropic triclinic representation must include shear response. The remaining out-of-plane components ($\sigma_{33},\sigma_{13},\sigma_{23}$) are excluded as they are structurally zero under plane stress for both representations.

Figure~\ref{fig:plane_stress_merged}(a)-(b) compares $\pi$-plane cross-sections at $\sigma_\text{m}=0$ within training coverage. The isotropic model accurately reproduces the target cross-section, whereas the triclinic model exhibits visible deviations for both specimens.

\paragraph{Extrapolation beyond training coverage.}
Beyond training coverage, the isotropic and triclinic models exhibit distinct extrapolation trends. Figure~\ref{fig:plane_stress_merged}(c)-(d) illustrates the yield loci of the learned models in the meridian plane across hydrostatic pressures from $-4500$ to $4500$~MPa, well beyond the coverage of the training paths derived from both the specimens. Note that the ground-truth yield surface remains open along the negative hydrostatic axis, indicating pressure-independent yield limits under compressive hydrostatic states. The triclinic models, trained separately on each specimen, form an artificial closed yield loop around the same hydrostatic pressures ($1100$-$1200$~MPa). Although the isotropic models also close at high pressures, their closure pressure scales with training coverage, and their extrapolation along the hydrostatic axis is substantially better than their triclinic counterparts'.

This improved extrapolation follows from the architecture's permutation invariance. Although training data lies only on the principal plane $s_3=0$, permutation invariance (Section~\ref{sec:symmetry}) enforces identical behavior on the untrained planes $s_1=0$ and $s_2=0$. The network's convexity then links this behavior smoothly across and beyond these three planes, regularizing the predicted yield surface even in regions never directly trained. The triclinic model is convex in its own coordinates too, but without permutation invariance, its convexity cannot transfer constraints to the untrained regions, where it extrapolates poorly.

\begin{figure}[t]
    \centering
    \includegraphics{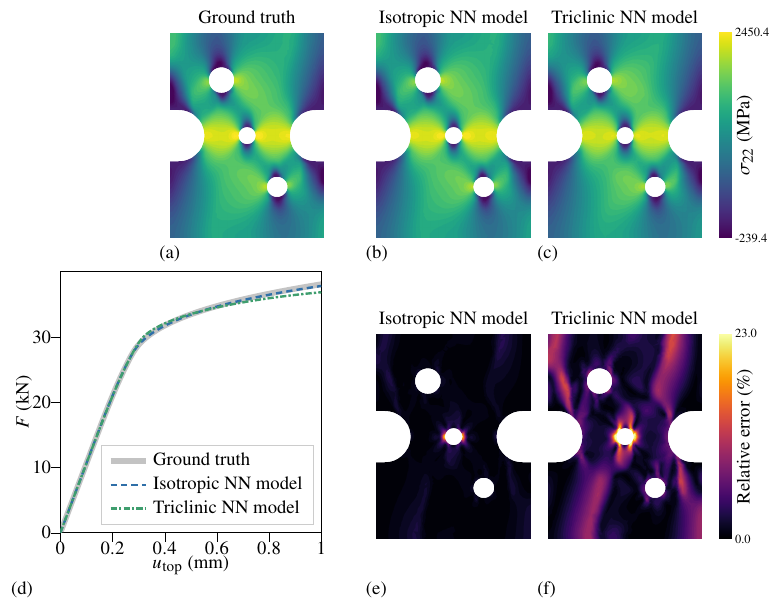}
    \caption{FEM validation on an unseen geometry. (a)-(c) Predicted $\sigma_{22}$ field: ground truth, isotropic NN model, triclinic NN model. (d) Force-displacement response (RMS relative error: 0.36\% isotropic vs.\ 1.66\% triclinic). (e)-(f) Relative error of the predicted $\sigma_{22}$ field, isotropic vs.\ triclinic (RMS relative error: 1.5\% vs.\ 5.3\%).}
    \label{fig:fem_validation}
\end{figure}

\paragraph{Finite element validation.}
To complement point-wise material evaluations, we test both models within full finite element boundary-value problems, where local constitutive errors can propagate through equilibrium calculations. For this assessment, we use the isotropic and triclinic models trained on the topology-optimized specimen, as they achieve the best performance in their respective categories. We evaluate both within a finite element solver on an unseen test geometry (Figure~\ref{fig:dogbone_to_geometry}c), and compare results against a reference solution obtained using the exact ground-truth LZY law on an identical mesh. The isotropic model achieves higher accuracy in both the force-displacement response and the predicted stress field. Figure~\ref{fig:fem_validation}(a)-(c) shows the predicted $\sigma_{22}$ stress fields alongside the ground truth. Figure~\ref{fig:fem_validation}(e)-(f) shows where that agreement actually comes from. The triclinic model's error spreads across the specimen domain, while the isotropic model's error remains localized near the central hole.

An additional study on the effect of isotropy enforcement, together with a comparison against a generic sequence-model baseline, is presented in Appendix~\ref{sec:material_patch_ablations}, using a full tensor dataset generated from a 3D material patch of the same underlying material law as in this study.

\subsection{Non-associativity in the open-cell foam RVE material}
\label{sec:axis4_flow_rule} 

\begin{figure}[t]
    \centering
    \begin{tabular}{>{\centering\arraybackslash}m{0.28\textwidth}@{\hspace{1.5em}}>{\centering\arraybackslash}m{5.5cm}@{\hspace{1.5em}}>{\centering\arraybackslash}m{5.5cm}}
    \includegraphics[width=0.28\textwidth]{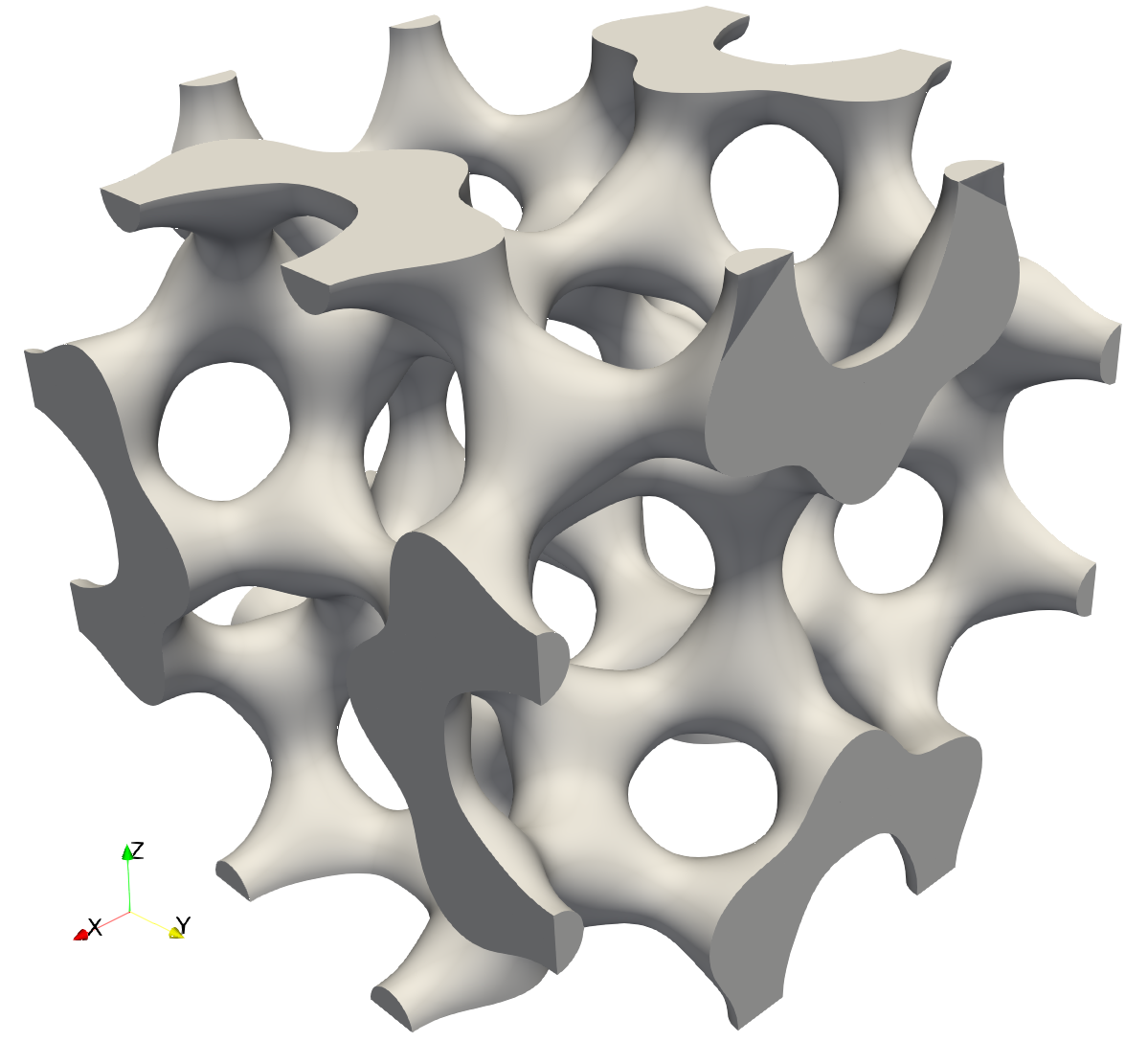} &
    \includegraphics[width=5.5cm]{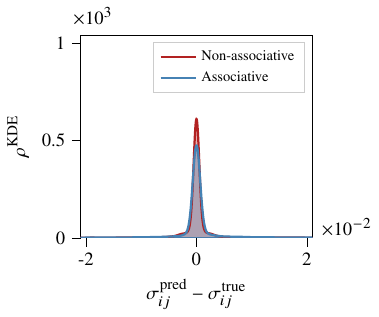} &
    \includegraphics[width=5.5cm]{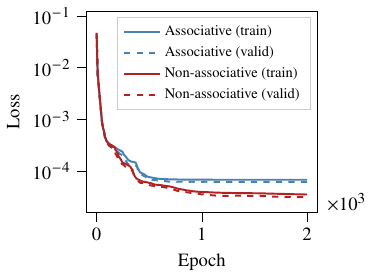} \\[-10pt]
    \makebox[\linewidth][l]{\small (a)} & \makebox[\linewidth][l]{\small (b)} & \makebox[\linewidth][l]{\small (c)} \\
    \end{tabular}
    \caption{(a) Periodic RVE geometry of the open-cell foam microstructure (relative density $\rho_\mathrm{rel}=0.2$), adapted from \citet{abendroth2024mode}. (b) Pooled signed residual distribution ($\sigma^\mathrm{pred}_{ij}-\sigma^\mathrm{true}_{ij}$), associative vs.\ non-associative. (c) Training and validation loss vs.\ epoch for both models.}
    \label{fig:foam8_geometry}
\end{figure}

While the preceding evaluations examine materials whose ground truth is a known classical law, we now apply the framework to an open-cell foam RVE dataset whose ground truth is not known in closed form. This dataset comprises homogenized stress-strain responses from the finite element simulation of a Weaire--Phelan foam structure RVE (Figure~\ref{fig:foam8_geometry}) subject to proportional stress loading \citep{abendroth2017ana,abendroth2024mode}. Although the microstructure-level material law is known (cf. Appendix~\ref{sec:source_rve}), the effective macroscopic behavior of the RVE, particularly the yield criterion, flow rule, and hardening law, is not known a priori.

We employ our thermodynamically consistent modeling framework to learn this effective behavior in two configurations. The first considers associative flow (Section~\ref{subsec:associated}); the second considers non-associative flow (Section~\ref{subsec:non-associated}), motivated by prior work reporting non-associative effective flow in open-cell foams \citep{abendroth2024mode}. Isotropic hardening alone was sufficient in both settings to capture the hardening behavior found in the foam RVE data.

Table~\ref{tab:axis4_summary} reports training and test metrics ($1-R^2$ and NRMSE) for both configurations. The non-associative configuration reaches a test-set NRMSE of 4.44\%, about 11\% lower, in relative terms, than the associative baseline (5.01\%). In Figure~\ref{fig:foam8_geometry}(b) we compare the pooled signed residual distributions of the two models. The non-associative model's residuals are more tightly concentrated near zero (standard deviation $0.00361$ vs.\ $0.00479$), consistent with its lower aggregate test NRMSE (Table~\ref{tab:axis4_summary}). We also see that both models converge smoothly over training (cf. Figure~\ref{fig:foam8_geometry}(c)).

\begin{figure}[t]
    \centering
    \includegraphics{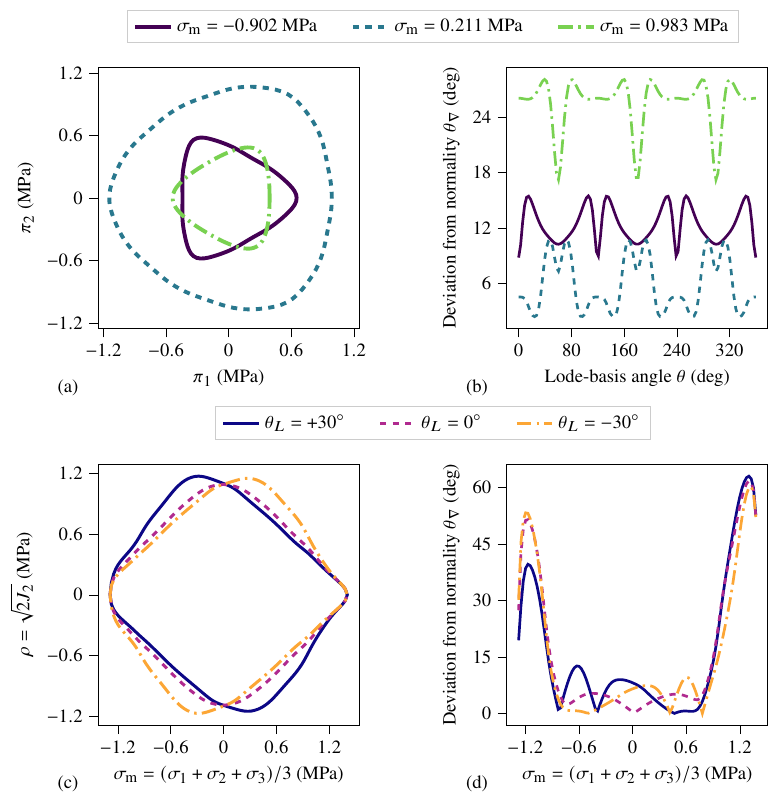}
    \caption{Non-associative model fitted to open-cell foam RVE data, showing the learned yield surface at a representative hardening level and its deviation from Hill's normality rule (angle between $\nabla\tilde\varphi$ and $\nabla\tilde\chi$; $0^\circ$ where locally associative). (a)-(b) $\pi$-plane cross-sections (a) and normality deviation vs.\ Lode angle (b), at three confining pressures $\sigma_\text{m}$. (c)-(d) Meridian cross-sections (c) and normality deviation vs.\ $\sigma_\text{m}$ (d), at three canonical Lode angles.}
    \label{fig:axis4_pi_plane}
\end{figure}

 As the difference in the two models' test-set NRMSE is small, we inspect whether the neural model in the non-associative configuration learns a flow potential $\tilde\chi$ that is meaningfully different from the yield function $\tilde\varphi$. In Figure~\ref{fig:axis4_pi_plane}, we visualize the learned yield surface geometry and quantify the flow non-associativity using the deviation between the yield-surface normal $\nabla\tilde\varphi$ and the flow-potential gradient $\nabla\tilde\chi$ evaluated at an intermediate hardening state. In the $\pi$-plane cross-sections (Figure~\ref{fig:axis4_pi_plane}(a)-(b)), the yield surface reaches its maximum area and circular profile near zero hydrostatic stress $\sigma_\text{m}$, contracting into a smaller, rotated triangular shape on the tension and compression sides. The three-fold symmetry of each individual cross-section follows directly from the model's enforced isotropy (permutation invariance over $(\sigma_1,\sigma_2,\sigma_3)$), whereas the rotation between the tension-side and compression-side triangles is an effect of the model's learned pressure sensitivity. The angular deviation from associativity is lower at intermediate hydrostatic pressures and increases toward pressure extremes (Figure~\ref{fig:axis4_pi_plane}(b),(d)), indicating more pronounced non-associative behavior under these conditions. Note that Figure~\ref{fig:axis4_pi_plane}(d) plots only one half of the meridian cross-section for each canonical Lode angle, as the other half is a mirror image due to the model's enforced isotropy.

\begin{figure}[t]
    \centering
    \begin{minipage}[t]{0.48\textwidth}
        \centering
        \resizebox{\linewidth}{!}{\includegraphics{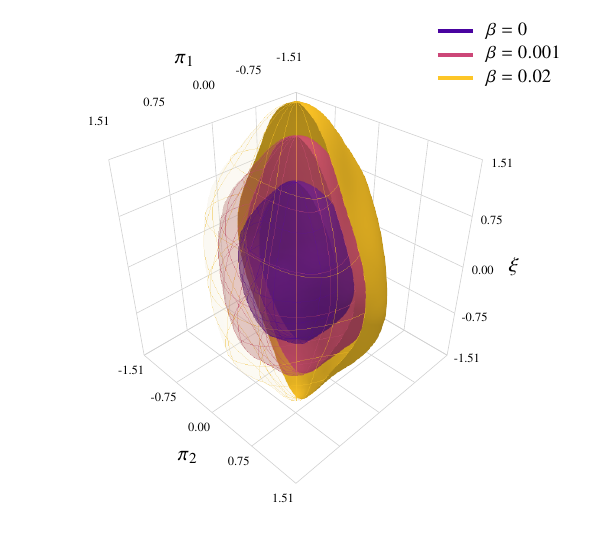}}\\[-10pt]
        \makebox[\linewidth][l]{\hspace{0.15\linewidth}\small (a)}
    \end{minipage}%
    \hfill
    \begin{minipage}[t]{0.48\textwidth}
        \centering
        \resizebox{\linewidth}{!}{\includegraphics{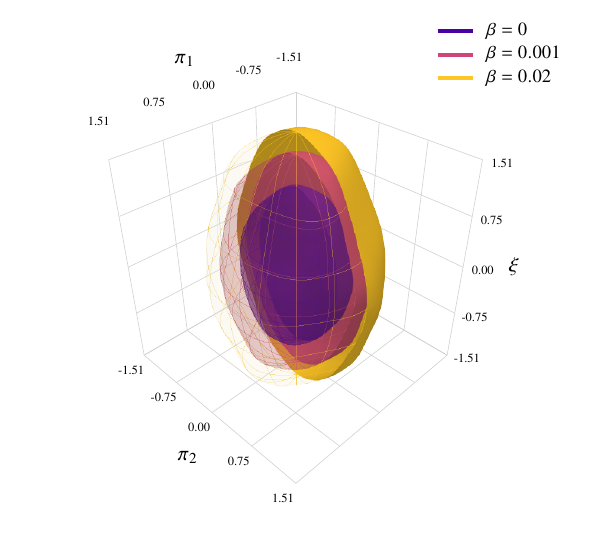}}\\[-10pt]
        \makebox[\linewidth][l]{\hspace{0.15\linewidth}\small (b)}
    \end{minipage}
    \caption{Non-associative model fitted to open-cell foam RVE data: (a) learned yield surface $\tilde\varphi=0$ and (b) flow potential $\tilde\chi$, full 3D view in Haigh--Westergaard coordinates ($\pi_1,\pi_2$: deviatoric plane; $\xi$: hydrostatic axis, vertical), at three isotropic-hardening levels $\beta$ (color, lightest to darkest = smallest to largest); each $\tilde\chi$ isosurface passes through the corresponding $\tilde\varphi=0$ boundary in (a). A cutting plane through the hydrostatic axis exposes the nested smaller-$\beta$ surfaces, with the removed half kept as a translucent shell so the full silhouette stays legible.}
    \label{fig:axis4_yield_surface_hw}
    \label{fig:axis4_flow_potential_hw}
\end{figure}

In Figure~\ref{fig:axis4_yield_surface_hw}, we visualize the yield surface $\tilde\varphi=0$ and flow potential $\tilde\chi$ in three-dimensional Haigh--Westergaard stress space across three hardening levels $\beta$. We remark that since we do not impose a degree-one homogeneity constraint on the flow potential or yield function, the models are free to learn subtle shape variations across hardening levels. Comparing the two potential surfaces directly illustrates the difference between $\tilde\chi$ isosurfaces and $\tilde\varphi=0$ boundaries, providing a visual representation of the non-associativity.

\begin{table}[t]
    \centering
    \small
    \caption{Performance metrics: associative vs.\ non-associative flow, open-cell foam RVE data.}
    \label{tab:axis4_summary}
    \begin{tabular}{lcccc}
        \toprule
        Model & Train $1{-}R^2$ & Test $1{-}R^2$ & Train NRMSE & Test NRMSE \\
        \midrule
        Associative & $1.39\times10^{-4}$ & $1.33\times10^{-4}$ & 3.39\% & 5.01\% \\
        Non-associative & $7.26\times10^{-5}$ & $7.51\times10^{-5}$ & 2.98\% & 4.44\% \\
        \bottomrule
    \end{tabular}
\end{table}

\section{Conclusion}
\label{sec:conclusion}

This work presented a neural plasticity framework that jointly learns a flexible yield function and its evolution laws within a rate-independent return-mapping scheme. The framework enforces material isotropy and thermodynamic consistency by construction across all learned components, remains numerically stable and trainable despite the eigenvalue-degeneracy singularities that arise in the return map's Newton--Raphson Jacobian, and can capture non-associative flow on complex material data.

Numerical evaluations substantiate these capabilities. The proposed architecture accurately reproduces classical constitutive laws across different materials, achieves substantial accuracy gains through isotropy enforcement, and captures complex microstructural responses in open-cell foam RVE data, including non-associative flow. The implicit differentiation formulation is degeneracy-safe, resolving gradient singularities at eigenvalue degeneracies and maintaining stability where alternative differentiation techniques lose accuracy or diverge (Appendix~\ref{app:eig_material_point_I1_I2}). These core principles---enforcing thermodynamic consistency, material symmetry, and convexity while resolving differentiability obstructions---extend to other constitutive classes built on a convex potential and solved implicitly via Newton iteration, such as rate-dependent (viscoplastic) and damage-coupled models. Such guarantees matter wherever learned models must remain thermodynamically admissible and numerically stable beyond their training distribution, as in metal forming, additive manufacturing, and geomechanical or structural design.

The present formulation addresses isotropic materials at infinitesimal strains. A natural direction for future work is extending the framework to anisotropic materials, such as rolled sheet metals or additively manufactured alloys, which requires replacing the spectral, permutation-invariant formulation with network representations tailored to lower material symmetry groups. Another is incorporating finite-strain kinematics and nonlinear hyperelastic free-energy functions to extend the framework to large-deformation plasticity regimes \citep{boes2026acco,jadoon2025auto}.

Since our simulation workflows are already compatible with ADiMU's data format \citep{ferreira2025auto}, an integration into the ADiMU framework would enable training directly on experimental displacement fields and boundary conditions from physical tests. Moreover, our compact, thermodynamically consistent architecture would give ADiMU a more efficient alternative to its current GRU surrogates (Appendix~\ref{sec:material_patch_ablations}), a benefit that could be inherited by specimen-design frameworks built on top of it, such as that of \citet{yin2026desi}. Finally, embedding this framework within a material-integrated design pipeline, following the topology-optimization approach demonstrated for hyperelastic neural networks \citep{vijayakumaran2025cons}, would move the learned model from an analysis tool to an integral part of structural design itself.

\section*{CRediT authorship contribution statement}

\textbf{Harikrishnan Vijayakumaran:} Conceptualization, Data curation, Formal analysis, Investigation, Methodology, Software, Validation, Visualization, Writing - original draft, Writing - review and editing.
\textbf{Shunyu Yin:} Data curation, Software, Validation, Writing - review and editing.
\textbf{Martin Abendroth:} Data curation, Writing - review and editing.
\textbf{Miguel A. Bessa:} Conceptualization, Funding acquisition, Methodology, Resources, Supervision, Writing - review and editing.
\textbf{Karl A. Kalina:} Conceptualization, Methodology, Supervision, Writing - original draft, Writing - review and editing.

\section*{Declaration of competing interest}
The authors declare that they have no known competing financial interests or personal relationships that could have appeared to influence the work reported in this paper.

\section*{Acknowledgements}

H.V. and M.A.B. gratefully acknowledge the financial support from the European Union's Horizon 2020 research and innovation program under the Marie Sk\l{}odowska-Curie grant agreement `LIGHTEN -- Ultralight membrane structures toward a sustainable environment' H2020-MSCA-ITN-2020-LIGHTEN-956547. M.A.B. also acknowledges that this effort was undertaken in part with the support from the Department of the Navy, Office of Naval Research, award number N000142612135. M.A. gratefully acknowledges the funding by the Deutsche Forschungsgemeinschaft (DFG, German Research Foundation) -- Project-ID 169148856 -- CRC 920: Multi-Functional Filters for Metal Melt Filtration -- A Contribution towards Zero Defect Materials. M.A. also acknowledges computing time on the Compute Cluster of the Faculty of Mathematics and Computer Science at Technische Universit\"at Bergakademie Freiberg, operated by the Computing Center (URZ) and funded by the Deutsche Forschungsgemeinschaft (DFG, German Research Foundation) -- Project-ID 397252409. K.A.K. thanks the DFG for the support within the Research Training Group GRK 2868 D${}^3$ -- Project Number 493401063.

\section*{Declaration of AI use}

The authors used a large language model (Claude, Anthropic) to improve the readability and refine the arguments of the manuscript text. All AI-assisted content was reviewed and edited by the authors, who take full responsibility for the content of this publication.

\appendix
\section{Notation}\label{app:notation}

Within this work, sets, tuples, and groups are denoted by script symbols, e.g., $\set S$, $\tuple T$, and $\tuple G$. Blackboard bold symbols are used for domains of numbers: $\mathbb Z$, $\mathbb N$, and $\mathbb R$ denote the integers, the natural numbers without zero, and the real numbers, respectively. A subscript attached to a number domain restricts it to elements satisfying the corresponding condition, e.g., $\mathbb{R}_{\geq 0}:=\left\{x\in\mathbb{R}\,|\,x \geq 0\right\}$.
Tensors of rank one and two are denoted by boldface italic and upright symbols, i.e., $\ve A, \ve B \in \Ln_1$ and $\te C, \te D \in \Ln_2$, respectively, where $\Ln_n$ denotes the space of real tensors of rank $n\in\mathbb N$ over the three-dimensional Euclidean space $\mathbb R^3$. Tensors of rank four are denoted by blackboard bold symbols, e.g., $\tttte A \in \Ln_4$.
Throughout this work, the Einstein summation convention is employed for tensor index notation unless stated otherwise.
Dyadic products of tensors are formed according to the order of the factors, e.g., $\ve A \otimes \ve B = A_k B_l \ve e_k \otimes \ve e_l$. Here, $\ve e_k,\ve e_l\in\mathbb R^3$ denote Cartesian basis vectors belonging to a Cartesian coordinate frame with fixed origin $O$, while $\otimes$ denotes the dyadic product. 
Single and double contractions of tensors are defined by $\ve A \cdot \te C = A_k C_{ki}\,\ve e_i$ and $\te C:\te D=C_{kl}D_{kl}$, respectively.
Transpose and inverse of a second-order tensor $\te C$ are denoted by $\te C^T$ and $\te C^{-1}$, respectively. Furthermore, $\tr[\boldsymbol{\bullet}]$, $\sym[\boldsymbol{\bullet}]$, $\det[\boldsymbol{\bullet}]$, and $\dev[\boldsymbol{\bullet}]$ denote the trace, symmetric part, determinant, and deviatoric part of a second-order tensor, respectively, e.g.\ $\tr[\te C]$ and $\dev[\te C]$.
The Euclidean norm of a vector and the Frobenius norm of a second-order tensor are defined by $|\ve A|:=\sqrt{A_iA_i}$ and $\|\te C\|:=\sqrt{C_{ij}C_{ij}}$, respectively. The set $\Sym:=\{\te S \in \Ln_2\,|\,\te S=\te S^T\}$ denotes the space of symmetric second-order tensors.
As is common in the literature, no explicit distinction is made between an algebraic structure and its underlying set. For example, the term group may refer either to the pair $(\set G,\cdot)$ or simply to its underlying set $\set G$, depending on the context.
The sets corresponding to the orthogonal group and the special orthogonal group are denoted by $\Orth:=\{\te A\in\Ln_2\,|\,\te A^T\cdot\te A=\ident\}$ and $\SO:=\{\te A\in\Ln_2\,|\,\te A^T\cdot\te A=\ident,\;\det\te A=1\}$, respectively. 
For the sake of readability, the arguments of functions are usually omitted throughout this work.
Moreover, no distinction is made between a function and its value; the same symbol is used for both whenever the meaning is clear from the context.

\section{Thermodynamic consistency}\label{app:thermo_consistency}

In this appendix, we demonstrate the thermodynamic consistency of models following the structure introduced in Section~\ref{sec:plasticity basics}. This is achieved by discussing the Clausius--Duhem inequality (CDI), $D:= - \dot \psi + \usig : \dot{\ueps}\ge 0$, which states that the dissipation rate $D$ must be non-negative for all admissible processes. To establish this, we follow the procedure developed by Coleman, Noll, and Gurtin \citep{coleman1963the,coleman1967ther}.

Applying the chain rule to $\psi(\ueps,\tuple I)$ gives
\begin{equation}
    D = \left(\usig - \diffp{\psi}{\ueps}\right):\dot{\ueps} - \diffp{\psi}{\tuple I}\bullet\dot{\tuple I}(\ueps,\lambda,\tuple I) \ge 0 \quad \forall\, \ueps,\dot\ueps\in\Sym,\,\tuple I\in\Sym\times\Sym\times\SR,\,\lambda\in\SR_{\ge0} \; \text{s.t. KKT} \; ,
\end{equation}
with $\bullet$ denoting the full contraction of all tuple elements. With the stress and force definitions of Eq.~\eqref{eq:thermo_forces}, the first term vanishes and $-\partial\psi/\partial\tuple I=\tuple F$, which reduces this to
\begin{equation}
    D = \tuple F(\ueps,\tuple I)\bullet\dot{\tuple I}(\ueps,\lambda,\tuple I) \ge 0 \quad \forall\, \ueps\in\Sym,\, \tuple I\in\Sym\times\Sym\times\SR,\, \lambda\in\SR_{\ge0} \; \text{s.t. KKT} \; .
    \label{eq:CDI_reduced}
\end{equation}

In both the non-associative and the associative case, the evolution law takes the form $\dot{\tuple I}=\lambda\,\partial\chi/\partial\tuple F$. For non-associative plasticity, this is Eq.~\eqref{eq:evolution:non_associated}; for associative plasticity, it follows from $\chi=\varphi+\sigma_{\text{y}_0}$ (Section~\ref{subsec:associated}), so that $\partial\chi/\partial\tuple F=\partial\varphi/\partial\tuple F$. Inserting it into Eq.~\eqref{eq:CDI_reduced} gives
\begin{equation}
    D = \lambda\,\tuple F \bullet \diffp{\chi(\tuple F)}{\tuple F} \; .
\end{equation}
For $\lambda=0$, $D=0$ follows trivially. For $\lambda>0$, the KKT condition $\lambda\varphi=0$ forces $\varphi(\tuple F)=0$, i.e., $\tuple F\in\partial\set E$. If $\chi$ is convex in $\tuple F$, the first-order convexity inequality, evaluated between $\tuple F$ and the force-free state, gives
\begin{equation}
\tuple F \bullet \diffp{\chi(\tuple F)}{\tuple F} \ge \chi(\tuple F) - \chi(\te 0,\te 0,0) \; ,
\label{eq:convexity_bound}
\end{equation}
so that $D\ge\lambda\left[\chi(\tuple F) - \chi(\te 0,\te 0,0)\right]$ for $\tuple F\in\partial\set E$. Non-negative dissipation therefore follows whenever
\begin{equation}
\chi(\tuple F) \ge \chi(\te 0,\te 0,0) \qquad \forall\, \tuple F\in\partial\set E \; .
\label{eq:general_condition}
\end{equation}
Since $\chi$ enters the evolution law only through its gradient, $\chi(\te 0,\te 0,0)$ can be chosen freely; with the normalization $\chi(\te 0,\te 0,0)=0$ of Section~\ref{subsec:evolution_thermo}, Eq.~\eqref{eq:general_condition} becomes Eq.~\eqref{eq:dissipation_condition_yield}. This condition restricts the value of $\chi$ on the yield surface only and does not require non-negativity of $\chi$ over all $\tuple F \in \Sym\times\Sym\times\SR$. Away from $\partial\set E$, $\lambda=0$ by the KKT conditions, so the sign of $\chi$ there does not enter the dissipation rate. It remains to verify Eq.~\eqref{eq:dissipation_condition_yield} for the specific parametrizations of the two cases.

\subsection{Non-associative plasticity}\label{app:TC_non_associated}

For the additive parametrization $\tilde\varphi=\tilde\chi+\Delta$ of Eq.~\eqref{eq:F_split}, Eq.~\eqref{eq:dissipation_condition_yield} can be guaranteed by a single constraint on $\Delta$. Since $\varphi=0$ on the yield surface, Eqs.~\eqref{eq:yield_function} and~\eqref{eq:F_split} give
\begin{equation}
    \tilde\chi(\usig^\text{eff}) + B = \sigma_{\text{y}_0} - \Delta(\usig^\text{eff}) \qquad \forall\, \tuple F \in \partial\set E \; .
\end{equation}
Inserting this into the extended flow potential of Eq.~\eqref{eq:flow_potential_extended} yields
\begin{equation}
\chi\big|_{\tuple F \in \partial \set E} = \sigma_{\text{y}_0}-\Delta(\usig^\text{eff}) + \chi^\text{kin}(\te A) + \chi^\text{iso}(B) \;\geq\; \sigma_{\text{y}_0}-\Delta(\usig^\text{eff}) \; ,
\label{eq:chi_at_yield}
\end{equation}
using $\chi^\text{kin},\chi^\text{iso}\geq0$ (Section~\ref{subsec:non-associated}). The right-hand side is strictly positive if
\begin{equation}
\Delta(\usig^\text{eff})<\sigma_{\text{y}_0} \qquad \forall\, \usig^\text{eff}\in\Sym \; ,
\label{eq:delta_bound}
\end{equation}
which we therefore impose as a constraint on the parametrization. Since the linear $B$ terms cancel between $\varphi$ and $\chi$, isotropic hardening requires no separate check. Moreover, with $\tilde\chi(\te 0)=0$, Eq.~\eqref{eq:delta_bound} gives $\varphi(\te 0,\te 0,0)=\Delta(\te 0)-\sigma_{\text{y}_0}<0$, so the force-free state lies in the elastic domain $\mathrm{int}(\set E)$.

This is a sufficient, not a necessary, route to Eq.~\eqref{eq:dissipation_condition_yield}. Only an upper bound on $\Delta$ is required, with $\Delta$ otherwise unconstrained and free to take arbitrarily large negative values where the material demands it. Since Eq.~\eqref{eq:chi_at_yield} holds for any yield surface the trained model produces, thermodynamic consistency is guaranteed for any setting of the weights. As $\tilde\varphi$ need not be convex, $\set E$ need not be convex either.

An independently parametrized $\tilde\varphi$, sharing no parameters or algebraic relationship with $\tilde\chi$, offers no comparable guarantee. Nothing ties $\partial\set E$ to the region where $\chi$ is non-negative, so Eq.~\eqref{eq:dissipation_condition_yield} would have to be checked empirically for each trained model.

\subsection{Associative plasticity}\label{app:TC_associated}
In the associative case, $\chi=\varphi+\sigma_{\text{y}_0}$ (Section~\ref{subsec:associated}), so $\chi$ is convex in $\tuple F$ exactly when $\varphi$ is. Since $\varphi=0$ on the yield surface, it follows that
\begin{equation}
    \chi(\tuple F) = \sigma_{\text{y}_0} \qquad \forall\, \tuple F \in \partial\set E \; .
\end{equation}
Eq.~\eqref{eq:dissipation_condition_yield} therefore reduces to
\begin{equation}
    \sigma_{\text{y}_0} \ge 0 \; , \qquad \text{i.e.,} \qquad \varphi(\te 0, \te 0, 0) = -\sigma_{\text{y}_0} \le 0 \; ,
\end{equation}
matching the condition of Section~\ref{subsec:associated}, and the dissipation rate satisfies $D \ge \lambda\sigma_{\text{y}_0} \ge 0$. Hence, thermodynamic consistency holds for any convex $\varphi$ and any non-negative initial yield stress $\sigma_{\text{y}_0}$.

\section{Input and output scaling}
\label{app:scaling}

Section~\ref{sec:scaling} outlines the general scaling conventions applied across the neural architecture in Section~\ref{sec:neural_architecture}. Physical-unit inputs are non-dimensionalized by a characteristic constant prior to network evaluation, and network outputs are subsequently rescaled by an equivalent physical factor. Consequently, each learned component's equations, as written, take and return quantities in physical units. This appendix specifies the characteristic scaling constants and origin corrections used for each component. Throughout this work, $\hat g$ denotes $\mathcal{S\!P}(\bar g^{\mathrm{NN}})$ for eigenvalue-based components and $\mathcal{S\!P}(g^{\mathrm{NN}})$ for purely scalar components, representing the (permutation-averaged, where applicable) neural network output composed with a softplus activation $\mathcal{S\!P}(\cdot)$ evaluated on scaled inputs. In Section~\ref{sec:neural_architecture}, $\hat g(\mathbf{x})$ serves as shorthand for $\hat g(\mathbf{x}/s)$, where $s$ denotes the characteristic scale of the component.

\paragraph{Scaling convention.}
Consider a component with argument $\mathbf{x}$, characteristic input scale $s$, output rescaling factor $r$, and origin correction operator $\mathrm{OC}\{\cdot\}$ (Section~\ref{sec:zero_states}). The non-dimensionalized formulation is expressed as
\begin{equation}
    f(\mathbf{x}) = r \cdot \mathrm{OC}\bigl\{\hat g(\mathbf{x}/s)\bigr\} \,,
    \label{eq:scaling_convention}
\end{equation}
where $\mathrm{OC}\{\cdot\}$ applies either a value correction ($\mathrm{OC}_\mathrm{val}\{\cdot\}$) or a combined value and gradient correction ($\mathrm{OC}_\mathrm{grad}\{\cdot\}$). In general, input and output scales share the same factor ($r = s$), with exceptions noted explicitly. This linear non-dimensionalization preserves the convexity, material symmetry, and origin behavior established in Section~\ref{sec:neural_architecture}.

\paragraph{Free-energy components (Section~\ref{sec:free_energy_architecture}).}
The free-energy potentials $\psi^\mathrm{iso}$ and $\psi^\mathrm{kin}$ operate on strain-like arguments ($\beta$ and the eigenvalues of $\ualpha$), yielding an input scaling constant of $\varepsilon_\text{s}$. Because free energies represent volumetric energy densities, their output scaling factor is the product $\sigma_\text{s}\varepsilon_\text{s}$:
\begin{equation}
    \psi^\mathrm{iso}(\beta) = \sigma_\text{s}\varepsilon_\text{s} \cdot \mathrm{OC}_\mathrm{grad}\bigl\{\hat g_{\psi^\mathrm{iso}}(\beta/\varepsilon_\text{s})\bigr\} \,, \qquad
    \psi^\mathrm{kin}(\ualpha) = \sigma_\text{s}\varepsilon_\text{s} \cdot \mathrm{OC}_\mathrm{grad}\bigl\{\hat g_{\psi^\mathrm{kin}}(\alpha_1/\varepsilon_\text{s},\alpha_2/\varepsilon_\text{s},\alpha_3/\varepsilon_\text{s})\bigr\} \,.
\end{equation}
Differentiating these energy functions with respect to strain-like arguments yields stress-like driving forces $B = -\partial\psi^\mathrm{iso}/\partial\beta$ and $\te A = -\partial\psi^\mathrm{kin}/\partial\ualpha$, matching physical stress dimensions.

\paragraph{Flow-potential and yield-function components (Sections~\ref{sec:flow_potential}--\ref{sec:yield_function}).}
The potentials $\tilde{\chi}$, $\chi^\mathrm{kin}$, and $\chi^\mathrm{iso}$ evaluate stress-like arguments comprising the principal eigenvalues of $\usig^\text{eff}$, $\te A$, and the scalar $B$. These functions use a stress scaling constant $\sigma_\text{s}$ for both input and output non-dimensionalization:
\begin{equation}
    \tilde{\chi}(\usig^\text{eff}) = \sigma_\text{s} \cdot \mathrm{OC}_\mathrm{val}\bigl\{\hat g_{\tilde\chi}(\sigma^\text{eff}_1/\sigma_\text{s},\sigma^\text{eff}_2/\sigma_\text{s},\sigma^\text{eff}_3/\sigma_\text{s})\bigr\} \,.
\end{equation}
The flow potential $\tilde\chi$ utilizes value correction $\mathrm{OC}_\mathrm{val}$ (Section~\ref{sec:flow_potential}), whereas $\chi^\mathrm{kin}$ and $\chi^\mathrm{iso}$ apply combined value and gradient corrections $\mathrm{OC}_\mathrm{grad}$.

The yield-surface correction network $g^{\mathrm{NN}}_\Delta$ (Eq.~\eqref{eq:F_delta}) is input-scaled by $\sigma_\text{s}$ using $u \coloneqq \mathrm{OC}_\mathrm{val}\{g^{\mathrm{NN}}_\Delta(\usig^\text{eff}/\sigma_\text{s})\}$, producing an output correction $\Delta = \sigma_{\text{y}_0}[1-\mathcal{S\!P}(c-u)]$ (Eq.~\eqref{eq:F_delta}). The network $g^{\mathrm{NN}}_\Delta$ is an unconstrained multilayer perceptron without an output activation function, allowing $\Delta$ to express signed variations. The output scaling parameter corresponds to the trainable initial yield stress $\sigma_{\text{y}_0} = \sigma_\text{s}\cdot\mathcal{S\!P}(\theta_{\text{y}_0})$, where $\theta_{\text{y}_0}$ is a trainable scalar parameter (Eq.~\eqref{eq:F_full}). This formulation enforces the physical bound $\Delta<\sigma_{\text{y}_0}$ (Section~\ref{sec:yield_function}) directly in physical stress units.

\section{Eigenvalue derivatives of a symmetric tensor}
\label{app:eig_derivatives}

Section~\ref{sec:eig_degeneracy} details the closed-form Hessian in Eq.~\eqref{eq:G_hessian_closed} and the divided-difference coefficient in Eq.~\eqref{eq:theta_branch} for $\tilde\chi(\sigma_1,\sigma_2,\sigma_3)$, following \citet{miehe1993comp} and \citet{souzaneto2008comp} (Appendix A). The same formulation applies directly to $\psi^\mathrm{kin}(\alpha_1,\alpha_2,\alpha_3)$ and $\chi^\mathrm{kin}(A_1,A_2,A_3)$ by replacing the stress tensor $\usig$ with $\ualpha$ or $\te A$.

\subsection{Lorentzian-regularized coefficient evaluation}
\label{app:eig_lorentzian}

For comparative evaluation in this appendix, we consider a Lorentzian-broadened coefficient $\vartheta_{ij}^{\mathrm{Lor}}$, commonly used in automatic differentiation of symmetric eigensolvers near degenerate eigenvalues \citep{liao2019diff}:
\begin{equation}
    \vartheta_{ij}^{\mathrm{Lor}} \coloneqq \left(\frac{\partial \tilde\chi}{\partial \sigma_i} - \frac{\partial \tilde\chi}{\partial \sigma_j}\right)\frac{\sigma_i - \sigma_j}{(\sigma_i - \sigma_j)^2 + \varepsilon^2}
    = \vartheta_{ij}\cdot\frac{\Delta_\sigma^2}{\Delta_\sigma^2+\varepsilon^2}, \qquad \Delta_\sigma \coloneqq \sigma_i - \sigma_j \,.
    \label{eq:theta_lorentzian}
\end{equation}
This formulation guarantees bounded values and evaluates to zero at exact coincidence. The relative deviation $\varepsilon^2/(\Delta_\sigma^2+\varepsilon^2)$ is localized within a small neighborhood of $\varepsilon$ near eigenvalue coincidence. Because the flow direction remains unaffected, the Newton iteration in Algorithm~\ref{ALG_RMA} retains quadratic convergence away from degeneracy. Within the $\varepsilon$-neighborhood of degenerate states, the update transitions to a quasi-Newton step. We evaluate this Lorentzian regularization solely for comparative purposes, while employing the exact branched formulation in Eq.~\eqref{eq:theta_branch} across all primary model evaluations.

\subsection{Material-point Hessian verification}
\label{app:eig_material_point_I1_I2}

\begin{figure}[tb]
    \centering
    \begin{tabular}{cc}
    \raisebox{-0.5\height}{\includegraphics{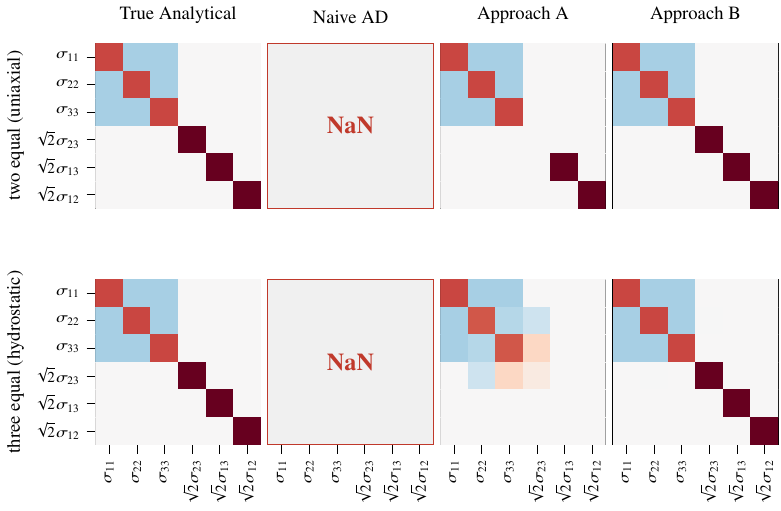}} & \raisebox{-0.5\height}{\includegraphics{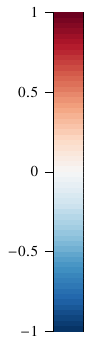}} \\
    \end{tabular}
    \caption{Stress Hessian of $J_2$ (Mandel $6\times 6$) across three stress states of increasing eigenvalue degeneracy (rows). Comparison includes the exact analytical Hessian (True Analytical), naive automatic differentiation through eigensolvers (Naive AD), Lorentzian-regularized evaluation (Approach A), and the exact closed-form construction from Section~\ref{sec:eig_degeneracy} (Approach B). Quantitative errors are presented in Table~\ref{tab:results_degeneracy}.}
    \label{fig:results_hessian_degeneracy}
\end{figure}

In Section~\ref{sec:eig_degeneracy}, we examine two treatments for handling eigenvalue-degeneracy singularities in the Newton Jacobian. Both formulations share the closed-form Hessian structure in Eq.~\eqref{eq:G_hessian_closed} and differ in their evaluation of $\vartheta_{ij}$ near coincidence. We compare the Lorentzian-regularized formulation (Approach A) and our exact L'H\^opital-branched formulation (Approach B) against naive automatic differentiation performed directly through the eigenvalue decomposition (Naive AD). Table~\ref{tab:results_degeneracy} compares these approaches against the exact analytical Hessian (True Analytical) for the representative invariant $J_2$ across three stress states: distinct eigenvalues, two coincident eigenvalues (uniaxial stress), and three coincident eigenvalues (hydrostatic stress). Error values reflect the relative Frobenius-norm deviation $\|\mathbf{H}-\mathbf{H}^\mathrm{true}\|/\|\mathbf{H}^\mathrm{true}\|$, where a value of zero indicates exact agreement.

\begin{table}[tb]
    \centering
    \small
    \caption{Relative Hessian error of $J_2$ with respect to stress.}
    \label{tab:results_degeneracy}
    \begin{tabular}{lccc}
        \toprule
        Degeneracy & Naive AD & Approach A & Approach B \\
        \midrule
        Distinct eigenvalues & exact & exact & exact \\
        Two coincident (uniaxial) & diverges & $0.447$ & exact \\
        Three coincident (hydrostatic) & diverges & $1.000$ & exact \\
        \bottomrule
    \end{tabular}
\end{table}

For distinct eigenvalues, all three numerical formulations match the analytical solution to machine precision. When two or three eigenvalues coincide, Naive AD diverges. Approach A remains bounded but introduces significant numerical error because the Lorentzian-regularized coefficient in Eq.~\eqref{eq:theta_lorentzian} evaluates to zero at coincidence, whereas the true limit in Eq.~\eqref{eq:theta_branch} is non-zero. Approach B maintains exact agreement across all evaluated stress states.

These numerical results confirm the failure modes predicted analytically in Section~\ref{sec:eig_degeneracy}. Naive AD suffers from accuracy loss and divergence under eigenvalue degeneracy. Approach A avoids divergence through bounded Jacobians but introduces approximation errors at degenerate states that decay over an $O((\varepsilon/\Delta)^2)$ neighborhood. Approach B yields exact derivatives across all degeneracy levels. Because this Hessian formulation relies solely on eigenvalue spectral decompositions, these results generalize directly to all eigenvalue-based scalar potentials evaluated in this work.

\section{Data generation}
\label{sec:data_generation}

The neural plasticity architecture described in Section~\ref{sec:neural_architecture} is calibrated using stress-strain path data. We distinguish two independent aspects of each dataset: the \emph{loading mechanism}, which dictates how a strain-stress path is generated (Section~\ref{sec:strain_path_generation}), and the \emph{constitutive source}, which determines the underlying physical response (Section~\ref{sec:data_sources}). As summarized in Table~\ref{tab:data_sources}, each dataset pairs a specific loading mechanism with a constitutive source.

To describe the physical information available for each constitutive source, we adopt three specific definitions. \emph{Synthetic} data is generated directly from closed-form constitutive laws (Section~\ref{sec:source_analytical}). \emph{RVE-simulated} data represents the homogenized macroscopic response computed from microscale finite element calculations where closed-form expressions are unavailable. \emph{Specimen-simulated} data consists of local Gauss-point stress-strain histories extracted from finite element simulations of loaded specimens, representing local field data obtained across a heterogeneous specimen. All datasets used in this work are generated numerically without physical testing.

\subsection{Loading mechanisms}
\label{sec:strain_path_generation}

\paragraph{Random load paths.}
Random strain paths are constructed by independently sampling and combining volumetric and deviatoric components. The peak volumetric strain per path is stratified uniformly across a specified range. The deviatoric loading direction is stratified uniformly around the Haigh--Westergaard $\pi$-plane. Its amplitude profile rises to a forward peak, reverses into the opposite direction up to a fraction $m$ of that peak, and returns to zero, with $m$ stratified across paths. This partial load reversal provides the strain signal required to distinguish isotropic from kinematic hardening. Paths generated via this mechanism operate purely in principal stress space with zero shear components. We use these random load paths in the model-expressivity evaluations in Section~\ref{sec:single_mechanism_studies}.

\begin{figure}[t]
    \centering
    \includegraphics{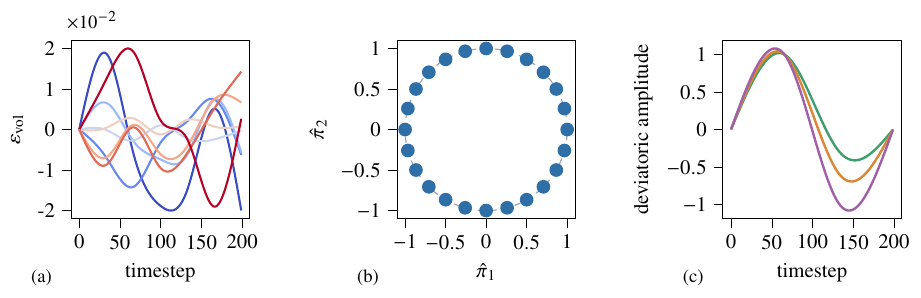}
    \caption{Stratified components of random load paths (Section~\ref{sec:strain_path_generation}). Left: volumetric peak strains distributed evenly across $[-\varepsilon_\mathrm{vol}^\mathrm{max},\varepsilon_\mathrm{vol}^\mathrm{max}]$, with color rank from lightest to darkest. Middle: deviatoric loading directions stratified uniformly in the Haigh--Westergaard $\pi$-plane on the $(\hat\pi_1,\hat\pi_2)$ basis, with a unit circle shown for reference. Right: forward and reverse deviatoric amplitude profiles shown for three values of the reversal fraction $m$.}
    \label{fig:strain_path_types}
\end{figure}

\paragraph{Proportional (radial) load paths.}
Proportional loading paths are generated in stress control by defining a fixed direction $\bar{\mathbf n}$ in principal stress space and prescribing stress as $\bar{\boldsymbol\sigma} = \lambda\,\bar{\mathbf n}$ for $\lambda \in [0, \lambda_\mathrm{max}]$. The scalar $\lambda_\mathrm{max}$ is set to reach a target equivalent plastic strain. This protocol produces monotonic, single-direction paths suitable for characterizing initial and evolving yield surface geometry under proportional loading. We apply this mechanism to generate the open-cell foam RVE dataset (Section~\ref{sec:data_sources}).

\paragraph{Specimen-level loadings.}
Specimen-level loading paths are extracted from full finite element simulations of loaded specimen geometries, including material patches, dogbones, and topology-optimized specimens. Under prescribed uniaxial displacement, geometric features induce heterogeneous, multiaxial, and non-proportional strain fields within the gauge region. We extract the local strain-stress time history at each Gauss point to form individual load paths. This procedure provides detailed local field data representative of measurements obtained from physical specimen testing. In these simulated cases, the underlying constitutive law is known and serves as a reference solution to evaluate model predictions. This mechanism generates the specimen-simulation datasets described in Section~\ref{sec:data_sources}.

\subsection{Constitutive sources}
\label{sec:data_sources}

Table~\ref{tab:data_sources} summarizes the datasets used in this work, categorized by constitutive source and loading mechanism.

\begin{table}[tb]
\centering
\small
\caption{Data sources used in this work.}
\label{tab:data_sources}
\begin{tabular}{p{5.5cm}p{3.7cm}p{5.5cm}}
\toprule
Source & Constitutive source & Loading mechanism \\
\midrule
1. Reference plasticity laws & Classical & Random (Section~\ref{sec:strain_path_generation}) \\[4pt]
2. Material patch & Specimen simulation & Specimen-level (Section~\ref{sec:strain_path_generation}) \\[4pt]
3. Dogbone/topology-optimized specimens & Specimen simulation & Specimen-level (Section~\ref{sec:strain_path_generation}) \\[4pt]
4. Open-cell foam RVE & RVE simulation & Proportional (Section~\ref{sec:strain_path_generation}) \\
\bottomrule
\end{tabular}
\end{table}

\subsubsection{Classical laws}
\label{sec:source_analytical}

\paragraph{Reference plasticity laws.}
We evaluate random load paths (Section~\ref{sec:strain_path_generation}) using closed-form expressions of ground-truth plasticity laws, defined by known free-energy functions, yield criteria, and flow potentials. These synthetic datasets verify that the framework recovers exact constitutive laws (Section~\ref{sec:single_mechanism_studies}). Each ground-truth law's expression represents a specific closed-form instance of the general formulation $\tilde\varphi(\usig^\text{eff}, B)$ and $\tilde\chi(\usig^\text{eff}, B)$ presented in Section~\ref{sec:neural_architecture}, where $\usig^\text{eff} = \usig + \te A$ and $B$ denotes the isotropic driving force. Tables~\ref{tab:analytical_yield_functions}, \ref{tab:analytical_iso_hardening_laws}, and \ref{tab:analytical_kin_hardening_laws} detail the yield criteria, isotropic hardening laws, and kinematic hardening laws used across evaluations. Table~\ref{tab:analytical_yield_params} lists the material parameters for each yield criterion, and Table~\ref{tab:analytical_hardening_params} provides the parameters for each hardening law.

\begin{table}[tb]
    \centering
    \small
    \caption{Material parameters for the yield criteria used in Section~\ref{sec:single_mechanism_studies}. Shared elastic parameters: $E_0=110\,000$~MPa, $\nu=0.33$.}
    \label{tab:analytical_yield_params}
    \begin{tabular}{ll}
        \toprule
        Law & Parameters \\
        \midrule
        Von Mises & $\sigma_{\text{y}_0}=900$~MPa \\
        Drucker--Prager (associative) & $\sigma_{\text{y}_0}=565$~MPa, $M_\varphi=0.05$, $M_\chi=0.05$ \\
        Drucker--Prager (non-associative) & $\sigma_{\text{y}_0}=565$~MPa, $M_\varphi=0.05$, $M_\chi=0.03$ \\
        LZY & $\sigma_{\text{y}_0}=900$~MPa, $a=1.0$, $b=0.05$, $c=1.0$, $d=0.5$ \\
        \bottomrule
    \end{tabular}
\end{table}

\begin{table}[tb]
    \centering
    \small
    \caption{Material parameters for the hardening laws used in Section~\ref{sec:single_mechanism_studies}, applied on a Von Mises base with $\sigma_{\text{y}_0}=900$~MPa (Table~\ref{tab:analytical_yield_params}).}
    \label{tab:analytical_hardening_params}
    \begin{tabular}{ll}
        \toprule
        Law & Parameters \\
        \midrule
        Isotropic, linear & $H=3000$~MPa \\
        Isotropic, power-law (Ludwik) & $s_1=700$~MPa, $s_2=0.5$ \\
        Isotropic, Voce & $Q=300$~MPa, $b=50$ \\
        Kinematic, linear (Prager) & $H=5500$~MPa \\
        Kinematic, Armstrong--Frederick & $H=5500$~MPa, $\gamma_\mathrm{af}=10$ \\
        \bottomrule
    \end{tabular}
\end{table}

\subsubsection{Specimen simulation}
\label{sec:source_fem}

\paragraph{Material patch.}
We simulate a material-patch specimen (Figure~\ref{fig:material_patch} in Appendix~\ref{app:material_patch_extra}) using finite element analysis following the model-updating methodology of ADiMU \citep{ferreira2025auto}. The simulation uses the LZY yield surface with power-law isotropic hardening, adopting the material parameters listed in Tables~\ref{tab:analytical_yield_params} and \ref{tab:analytical_hardening_params}. This dataset is evaluated in Appendix~\ref{sec:material_patch_ablations}.

\paragraph{Dogbone and topology-optimized specimens.}
A second specimen-simulation dataset includes standard dogbone and topology-optimized geometries (Figure~\ref{fig:dogbone_to_geometry}), generated following the specimen design methodology of \citet{yin2026desi}. The underlying ground-truth material law uses the LZY yield surface and power-law isotropic hardening parameters specified in Tables~\ref{tab:analytical_yield_params} and \ref{tab:analytical_hardening_params}. We use this dataset in the plane-stress symmetry evaluations in Section~\ref{sec:plane_stress_symmetry_ablation}.

\subsubsection{RVE simulation}
\label{sec:source_rve}

\paragraph{Open-cell foam RVE dataset.}
We evaluate a periodic representative volume element (RVE) modeling an open-cell ceramic foam microstructure (Figure~\ref{fig:foam8_geometry} in Section~\ref{sec:axis4_flow_rule}). The microstructure has a relative density of $\rho_\mathrm{rel}=0.2$, with matrix material properties $E=10\,000$~MPa and $\nu=0.3$. At the microscale, the solid phase follows isotropic Von Mises plasticity with linear hardening ($\sigma_{y_0}=20$~MPa, $H=10$~MPa, Table~\ref{tab:analytical_iso_hardening_laws}). The RVE is subjected to periodic boundary conditions under proportional stress paths, generating data in support of metal-melt filtration filter design \citep{abendroth2017ana, abendroth2024mode}. We utilize this dataset in Section~\ref{sec:axis4_flow_rule}.

\section{Additional results on material-patch data}
\label{app:material_patch_extra}

This section extends the plane-stress isotropy-enforcement study of Section~\ref{sec:plane_stress_symmetry_ablation} to a full three-dimensional stress state, using the material-patch dataset described in Appendix~\ref{sec:source_fem}. Figure~\ref{fig:material_patch} illustrates the specimen setup.

\begin{figure}[tb]
    \centering
    \begin{tabular}{c@{\hspace{2em}}c}
    \includegraphics{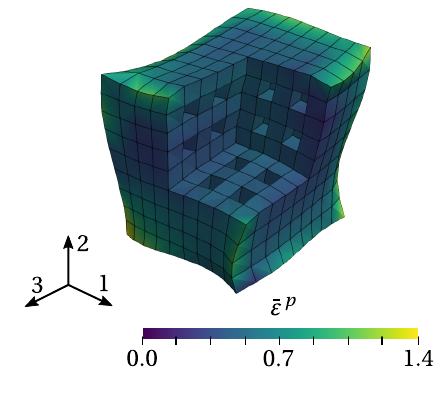} &
    \includegraphics{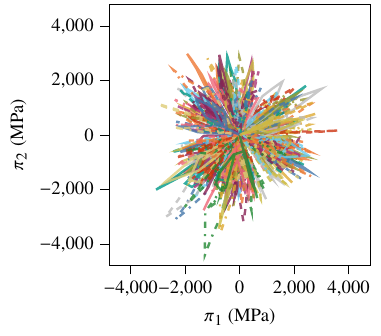} \\[-10pt]
    \makebox[212pt][l]{\small (a)} & \makebox[182.449pt][l]{\small (b)} \\
    \end{tabular}
    \caption{Specimen geometry and loading configuration for the material-patch dataset. (a) Accumulated plastic strain field $\bar\varepsilon^{\mathrm{p}}$ at the final loading step. (b) $\pi$-plane projection of the 256 training load paths used in the $n^\text{load}=256$ sample-efficiency evaluation (Section~\ref{sec:material_patch_ablations}), diagonalized to principal stresses.}
    \label{fig:material_patch}
\end{figure}

\subsection{Sample efficiency and symmetry-prior ablation on material-patch data}
\label{sec:material_patch_ablations}

\begin{figure}[tb]
    \centering
    \includegraphics{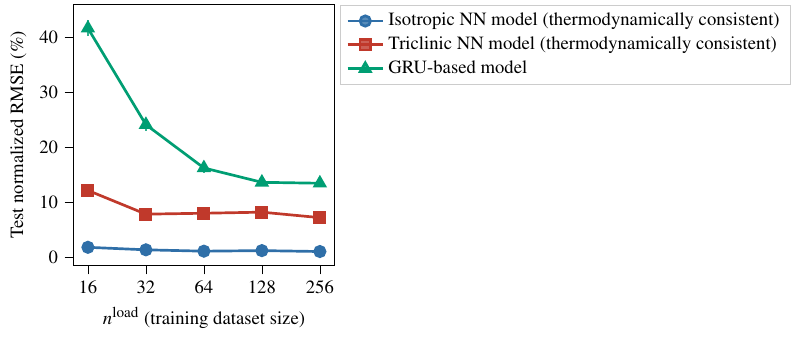}
    \caption{Seed-averaged (3 seeds) test normalized RMSE vs.\ training dataset size for material-patch LZY data: isotropic NN model (this work) vs.\ triclinic NN model vs.\ generic sequence model (GRU). Markers indicate seed means, and vertical error bars show minimum and maximum values across seeds.}
    \label{fig:material_patch_nrmse_vs_n}
\end{figure}

We conduct two additional evaluations on this dataset using a unified experimental design. We train three model formulations under identical conditions (elastic parameters, isotropic-hardening mechanisms, and optimization settings) across nested dataset sizes from $n^\text{load}=16$ to $256$ paths, evaluating each against a separate held-out test set of 511 paths. To account for optimization variability across training runs, we train three random seeds for every dataset size and model formulation. Table~\ref{tab:material_patch_nrmse_summary} and Figure~\ref{fig:material_patch_nrmse_vs_n} report seed-averaged test errors and range bounds.

\paragraph{Architecture comparison.}
We test whether this physics-based structure provides a genuine benefit by comparing it against a generic sequence model with no physics built in. The baseline architecture is a stacked recurrent neural network (GRU) that predicts plastic-strain increments directly from strain increments and elastic trial stresses, without enforcing yield criteria, convexity, or consistency conditions. Following \citet{ferreira2025auto}, the GRU model uses 3 hidden layers with 444 units, totaling approximately 3 million trainable parameters. This is far more than the 673 trainable parameters combined across our flow potential $\tilde\chi$ (Section~\ref{sec:flow_potential}) and isotropic-hardening free energy $\psi^\mathrm{iso}$ (Section~\ref{sec:free_energy_architecture}, Eq.~\eqref{eq:psi_iso_integral}), so its higher error, despite this larger parameter count, reflects the lack of physics-based structure.

Across all dataset sizes, our physics-constrained model maintains a seed-averaged test NRMSE between 1.04\% and 1.81\% (Table~\ref{tab:material_patch_nrmse_summary}, Figure~\ref{fig:material_patch_nrmse_vs_n}), whereas the GRU model's test error remains higher throughout, reaching 13.47\% NRMSE at $n^\text{load}=256$. Seed averaging confirms a monotonic reduction in test error with dataset size across all evaluated models.

\begin{table}[tb]
    \centering
    \small
    \caption{Seed-averaged (3 seeds) test NRMSE vs.\ training dataset size for material-patch LZY data: isotropic NN model (this work) vs.\ triclinic NN model vs.\ generic sequence model (GRU). The ratio indicates triclinic NRMSE divided by isotropic NRMSE.}
    \label{tab:material_patch_nrmse_summary}
    \begin{tabular}{c ccc c}
        \toprule
        $n^\text{load}$ & Isotropic & Triclinic & GRU & Ratio \\
        \midrule
        16 & 1.81\,$\pm$\,0.17\% & 12.17\,$\pm$\,0.43\% & 41.68\,$\pm$\,1.77\% & 6.7$\times$ \\
        32 & 1.35\,$\pm$\,0.31\% & 7.85\,$\pm$\,0.41\% & 24.14\,$\pm$\,1.35\% & 5.8$\times$ \\
        64 & 1.11\,$\pm$\,0.02\% & 8.02\,$\pm$\,0.39\% & 16.25\,$\pm$\,1.04\% & 7.2$\times$ \\
        128 & 1.21\,$\pm$\,0.18\% & 8.21\,$\pm$\,0.61\% & 13.64\,$\pm$\,0.45\% & 6.8$\times$ \\
        256 & 1.04\,$\pm$\,0.16\% & 7.21\,$\pm$\,1.09\% & 13.47\,$\pm$\,0.31\% & 6.9$\times$ \\
        \bottomrule
    \end{tabular}
\end{table}

\paragraph{Symmetry-prior comparison.}
We isolate the effect of enforcing material isotropy (Section~\ref{sec:symmetry}) by comparing our permutation-invariant spectral yield representation against a triclinic baseline with no symmetry enforced. The triclinic baseline shares identical layer widths and depths but operates directly on the components of the Cauchy stress $\usig$ in Mandel notation (Remark~\ref{remark:triclinic}). Because the triclinic function space strictly encompasses isotropic functions, the network must learn material symmetries entirely from data.

Our isotropy-enforced model achieves 5.8 to 7.2 times lower seed-averaged test NRMSE than the triclinic representation across all dataset sizes (Table~\ref{tab:material_patch_nrmse_summary}). As shown in Figure~\ref{fig:material_patch_nrmse_vs_n}, increasing training data from $n^\text{load}=16$ to $64$ reduces test NRMSE for the isotropic model, after which error plateaus near 1.04--1.21\%. The triclinic baseline plateaus at approximately 7--8\% NRMSE beyond $n^\text{load}=32$.

\begin{figure}[tb]
    \centering
    \includegraphics{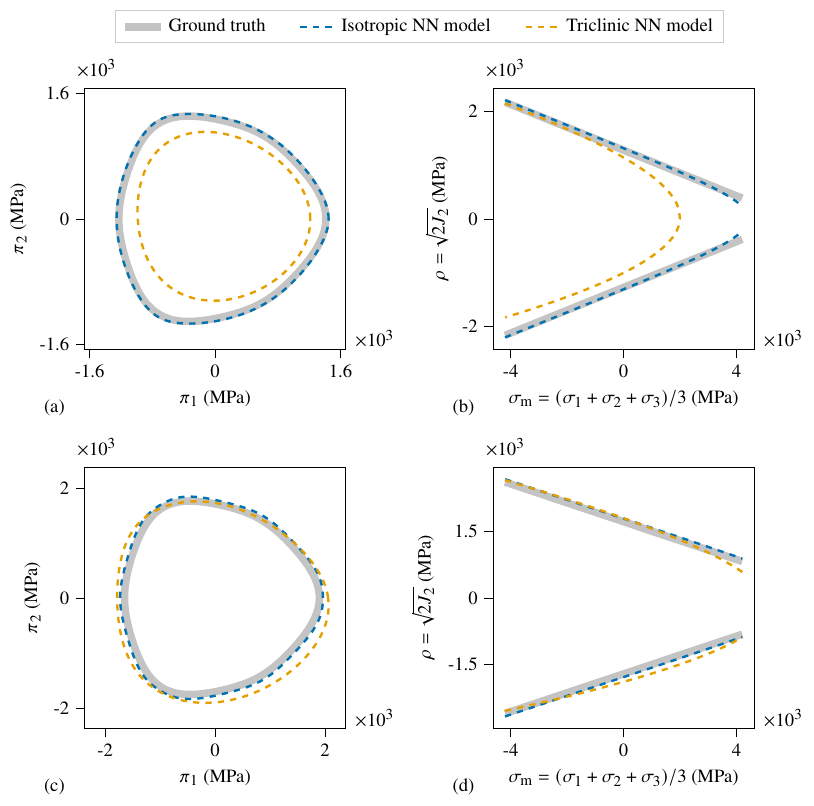}
    \caption{$\pi$-plane cross-sections (a, c) and meridian cross-sections (b, d) comparing the isotropic NN model, triclinic NN model, and ground-truth LZY law for $n^\text{load}=16$ material-patch data. Panels (a, b) show initial yield surfaces ($\beta=0$), and panels (c, d) compare yield surfaces at a matched high-hardening level ($\bar\varepsilon^{\mathrm{p}}\approx0.194$). The ground truth is shown as a semi-transparent band, with learned model predictions plotted as dashed lines.}
    \label{fig:no_isotropy_shape}
\end{figure}

Figure~\ref{fig:no_isotropy_shape} compares learned yield surfaces against the ground-truth LZY law with power-law isotropic hardening (Tables~\ref{tab:analytical_yield_params}, \ref{tab:analytical_hardening_params}), which fits the numerical dataset within 3.29\% NRMSE. Because the two representations parameterize hardening differently, we evaluate cross-sections at a matched physical hardening state, the 90th percentile of test-set plastic strain ($\bar\varepsilon^{\mathrm{p}}\approx0.194$), rather than a shared value of each model's own hardening variable $\beta$. At this hardening level (Figure~\ref{fig:no_isotropy_shape}c, d), both models predict larger yield loci than the ground truth.

\bibliographystyle{plainnat}
\bibliography{lit}

\end{document}